\documentclass[fleqn,usenatbib]{mnras}

\usepackage{newtxtext,newtxmath}
\usepackage{siunitx}

\usepackage[T1]{fontenc}
\usepackage[utf8]{inputenc}

\DeclareRobustCommand{\VAN}[3]{#2}
\let\VANthebibliography\thebibliography
\def\thebibliography{\DeclareRobustCommand{\VAN}[3]{##3}\VANthebibliography}

\usepackage{graphicx}	
\usepackage{amsmath}	
\usepackage[utf8]{inputenc}

\newcommand{\HI}{H\,\textsc{i}}
\newcommand{\Lya}{Ly$\alpha$}

\title[Uncertainty-Aware CNN for Ly$\alpha$ Forest]{Uncertainty-Aware Deep Learning for the Ly$\alpha$ Forest: CNN-Based Absorber Detection and Characterization}

\author[]{
Paryag Sharma,$^{1}$\thanks{panditparyag@gmail.com}
Vikram Khaire,$^{2}$
Ting-Yun Cheng,$^{3}$\thanks{t.cheng@rug.nl}
Hum Chand,$^{1}$
Prakash Gaikwad$^{4}$
\\
$^{1}$Department of Physics and Astronomical Science, Central University of Himachal Pradesh, Dharamshala, 176215, India\\
$^{2}$Department of Physics, Indian Institute of Technology Tirupati, Tirupati, Andhra Pradesh 517619, India\\
$^{3}$ Kapteyn Astronomical Institute, University of Groningen, Landleven 12 (Kapteynborg, 5419), 9747 AD Groningen, The Netherlands\\
$^{4}$ Department of Astronomy, Astrophysics and Space Engineering, Indian Institute of Technology Indore, Simrol, MP 453552, India
}

\date{Accepted XXX. Received YYY; in original form ZZZ}

\pubyear{\the\year{}}

\begin{document}
\label{firstpage}
\pagerange{\pageref{firstpage}--\pageref{lastpage}}
\maketitle

\begin{abstract}
The Ly$\alpha$ forest is a powerful probe of the intergalactic medium and small-scale matter distribution, but deriving absorber properties traditionally requires computationally expensive Voigt-profile fitting. We present a convolutional neural network (CNN) that identifies and characterizes H~\textsc{i} Ly$\alpha$ absorbers directly from quasar spectra. The model is trained on synthetic spectra generated from the IllustrisTNG simulation and fitted with the \textsc{viper} Voigt-profile fitting code to provide training labels. The network simultaneously predicts absorber presence, column density ($N_{\rm HI}$), Doppler parameter ($b_{\rm HI}$), and line centroid.
On simulated spectra, the CNN reproduces the \textsc{viper} reference catalogue with an F1 score of $\sim0.8$, with mean absolute errors of $\sim0.18$ in $\log N_{\rm HI}$ and $\sim0.10$ in $\log b_{\rm HI}$. It reproduces the H~\textsc{i} column density distribution function (CDDF) and the $b_{\rm HI}$--$N_{\rm HI}$ relation obtained from \textsc{viper}, recovering similar CDDF slopes and a lower-envelope relation with an RMS difference of only $0.36~{\rm km~s^{-1}}$. Applied to high-resolution UVES spectra, performance decreases to an F1 score of $\sim0.5$, with mean absolute errors of $\sim0.35$ in $\log N_{\rm HI}$ and $\sim0.22$ in $\log b_{\rm HI}$. Latent-space analysis reveals a significant domain shift between the simulated and observational spectra, contributing to the reduced performance. Nevertheless, the CNN reproduces the observed CDDF and $b_{\rm HI}$--$N_{\rm HI}$ distributions obtained from the \textsc{viper} reference catalogue, yielding CDDF slopes consistent with \textsc{viper} and a lower-envelope RMS difference of $3.20~{\rm km~s^{-1}}$ relative to \textsc{viper}. Monte Carlo dropout is implemented during inference to quantify predictive uncertainties. Together with its computational efficiency, the method provides a scalable and uncertainty-aware framework for Ly$\alpha$ forest analysis in upcoming spectroscopic surveys.

\end{abstract}




\section{Introduction}
A significant fraction of the baryons in the Universe resides in the tenuous regions between galaxies, collectively known as the intergalactic medium (IGM), characterized by very low densities $n \sim 10^{-7}~\mathrm{cm^{-3}}$ and temperatures spanning $T \sim 10^{4} - 10^{7}~\mathrm{K}$ depending on phase \citep{Meiksin2009RvMP...81.1405M}.
One of the most luminous classes of objects in the Universe, quasars are powered by accretion onto supermassive black holes. As their radiation traverses cosmological distances, it passes through numerous intervening \HI\ structures that absorb light at the redshifted Lyman-$\alpha$ (\Lya) resonance line (rest wavelength 1215.67~\AA), imprinting a dense sequence of absorption features on the quasar continuum. 
This ensemble of absorption lines, called the \Lya\ forest, provides one of the most sensitive observational probes of the diffuse IGM \citep{Lynds1971ApJ...164L..73L,Rauch1998ARA&A..36..267R}. These absorption signatures trace fluctuations in the underlying matter distribution on scales from tens of kiloparsecs to several megaparsecs \citep[e.g.][]{Cristiani1995MNRAS.273.1016C, Fang1996ApJ...462...77F}. 
Because neutral hydrogen in the low-density IGM is highly sensitive to the 
temperature–density relation 
and the ionizing ultraviolet background, the \Lya\ forest provides a detailed probe of the thermal state of diffuse gas in the IGM
\citep[][]{Schaye2001ApJ...559..507S, Bolton2008MNRAS.386.1131B, Becker2013MNRAS.430.2067B, Gaikwad2021MNRAS.506.4389G, Hu2025MNRAS.536....1H,Hu2026arXiv260605006H} 
and hydrogen photoionization rate \citep{Becker2013MNRAS.436.1023B, Khaire2019MNRAS.486..769K}  and is used to calibrate ultraviolet background models \citep{ Haardt2012ApJ...746..125H,Khaire2015ApJ...805...33K,Khaire2019MNRAS.484.4174K,Puchwein2019MNRAS.485...47P,Faucher2020MNRAS.493.1614F}.
The forest’s flux statistics further reflect the amplitude and shape of the small-scale matter power spectrum, enabling constraints on cosmological models and on the nature of dark matter \citep[][]{Viel2013MNRAS.429.1734V,Irsic2017PhRvD..96b3522I,Rogers2021PhRvL.126g1302R}. Precise measurements of the \Lya\ forest have been used to constrain the growth of structure and the expansion history of the Universe through their sensitivity to the large-scale clustering of matter, including baryon acoustic oscillation (BAO) measurements and flux power spectrum analyses \citep[][]{Busca2013A&A...552A..96B, Slosar2013JCAP...04..026S, Bourboux2020ApJ...901..153D, Alam2021PhRvD.103h3533A,Pedersen2023ApJ...944..223P}. 
Together, these properties make the \Lya\ forest a powerful astrophysical and cosmological probe across a broad range of redshifts.

Extracting physical information from the \Lya\ forest, however, is challenging. One approach 
involves decomposing each absorption feature into a Voigt profile, from which one can infer the column density ($N_{\rm HI}$), Doppler parameter ($b_{\rm HI}$), and absorber redshift. Although widely used tools such as \textsc{vpfit} \citep{Carswell2014ascl.soft08015C}, \textsc{viper} \citep{Gaikwad2017MNRAS.467.3172G}, \textsc{BayesVP} \citep{Liang2017arXiv171009852L}, \textsc{GVPFIT} \citep{Bainbridge2017MNRAS.468.1639B}, and \textsc{VoigtFit} \citep{Krogager2018arXiv180301187K} have become sophisticated and robust, Voigt-profile fitting remains computationally intensive and sensitive to model initialization. In particular, Voigt-profile fitting tools such as \textsc{vpfit} and \textsc{viper} have been successfully applied to large Ly$\alpha$ forest datasets to characterize the thermal state of the intergalactic medium, measure the H~\textsc{i} absorber population, and constrain the ionizing ultraviolet background from high-resolution quasar spectra \citep{Rudie2012ApJ...757L..30R,Gaikwad2017MNRAS.466..838G,Gaikwad2020MNRAS.494.5091G,Gaikwad2021MNRAS.506.4389G}. These studies demonstrate the scientific power of automated and semi-automated Voigt-profile fitting while also emphasizing the need for faster analysis techniques for future large spectroscopic surveys.
In practice, full forests often contain blended features, wavelength-dependent noise, and instrumental systematics that require significant human intervention to resolve. 
As a result, scaling traditional fitting methods to the millions of quasar spectra expected from upcoming surveys such as DESI \citep{DESI2025arXiv250314745D} is increasingly impractical.

Machine learning (ML) offers a promising path forward. Deep neural networks, particularly convolutional neural networks (CNNs), have already demonstrated strong performance across a wide range of astronomical applications, including galaxy morphology classification \citep{Cheng2020MNRAS.493.4209C,Cheng2021MNRAS.507.4425C,Walmsley2022MNRAS.509.3966W}, strong lens modeling \citep{Metcalf2019A&A...625A.119M}, transient identification \citep{Muthukrishna2019PASP..131k8002M}, and a variety of spectroscopic tasks. 
Within absorption-line and spectroscopic studies, ML techniques have been used for detecting damped \Lya\ systems \citep{Garnett2017MNRAS.472.1850G,Parks2018MNRAS.476.1151P}, 
recovering the \Lya\ optical depth field \citep{Huang2021MNRAS.506.5212H}, 
inferring the matter power spectrum \citep{Veiga2021arXiv210709082V}, and characterizing metal-line systems such as Mg\,{\sc ii} \citep{Stemock2024AJ....167..287S}. 
More generally, ML has been applied to continuum normalization \citep{Pistis2025A&A...698A.292P}, spectral classification in large surveys \citep{Sharma2020MNRAS.491.2280S}, and the emulation of high-resolution spectral models \citep{Cabayol-Garcia2023MNRAS.525.3499C}. Together, these efforts highlight the growing role of ML as a flexible and scalable framework for interpreting complex spectroscopic datasets.

Two recent approaches are particularly relevant to the present work. 
\citet{Cheng2022MNRAS.517..755C} train their CNN using fully synthetic Ly$\alpha$-forest spectra generated with the \textsc{pyigm}\footnote{\url{https://github.com/pyigm/pyigm}} package. 
In their framework, each mock spectrum corresponds to a quasar at $z_{qso}=3$ and contains a large number of Ly$\alpha$ absorption systems drawn from the H\,\textsc{i} column-density distribution function \citep{Prochaska2014MNRAS.438..476P}, with Doppler parameters sampled from the \cite{Hui1999ApJ...517..541H} distribution and
redshifts drawn from the redshift-dependent incidence rate $l(z) \equiv dN/dz$.
These absorption features are rendered as Voigt profiles, convolved with a high-resolution instrumental kernel and degraded to different signal-to-noise ratios.  
Each pixel in the simulated forest is assigned “ground-truth’’ labels and the CNN is trained using a sliding-window that moves across the entire spectrum one pixel at a time, predicting physical parameters for the central pixel of each window. 
By operating directly on full simulated spectra containing numerous and blended features, their model locates and characterizes absorbers without requiring prior knowledge of line positions.

In contrast, \citet{Jalan2024A&A...688A.126J} construct their training data not from full spectra, but from isolated Voigt-profile absorption features designed to mimic low-redshift Ly$\alpha$ lines observed with HST/COS \citep{Danforth2016ApJ...817..111D}. 
Individual Voigt profiles are generated by randomly sampling physical parameters from uniform distributions spanning the parameter ranges observed in the HST/COS data. The simulated profiles are then convolved with the COS line-spread function, rebinned to the COS pixel scale, and degraded to a randomly selected SNR.  
Each synthetic absorption line or blend of two components for double systems is then placed at the centre of a fixed 301-pixel window, with the surrounding pixels padded by continuum and Gaussian noise. 
Because the simulated profiles are generated from known input parameters, the labels used for training are directly given by the injected absorber properties, including the number of components, Doppler parameters, and H\,\textsc{i} column densities. Their architecture consists of a classifier that distinguishes single from double components, followed by dedicated regression networks that infer $b$ and $N_{\rm HI}$ for the identified components. To assess performance on observational data, the predictions are later compared with absorber catalogues produced using traditional Voigt-profile fitting approaches such as \textsc{viper} and \cite{Danforth2016ApJ...817..111D}. Since every training example is centred on a known absorber, this approach assumes that line locations are identified beforehand and focuses on accurate parameter recovery rather than automated line finding.

In this study, we build upon complementary aspects of both \citet{Cheng2022MNRAS.517..755C} and \citet{Jalan2024A&A...688A.126J}. Whereas \citet{Cheng2022MNRAS.517..755C} generate their training spectra using the analytic prescriptions implemented in \textsc{pyigm}, we instead construct our training set directly from the IllustrisTNG cosmological simulation \citep{Nelson2019ComAC...6....2N}. This provides a physically self-consistent environment in which gas kinematics, density structure, and metal enrichment emerge naturally from cosmological hydrodynamics and feedback, allowing the network to learn from spectra that better capture the complexity of realistic absorption structures. In contrast, \citet{Jalan2024A&A...688A.126J} demonstrated that machine-learning methods trained on physically motivated synthetic absorption systems can effectively recover absorber properties and validated their predictions against traditional Voigt-profile fitting approaches, including \textsc{viper}.
Motivated by these studies, we combine physically realistic synthetic spectra with absorber labels derived from the \textsc{viper} Voigt-profile fitting code, ensuring consistency with the methodology commonly used for observational analyses. Unlike the fixed, line-centred windows adopted by \citet{Jalan2024A&A...688A.126J}, however, we follow the sliding-window strategy of \citet{Cheng2022MNRAS.517..755C}, enabling the network to scan full spectra and simultaneously learn absorber identification and parameter estimation. We evaluate the resulting model on both simulated and observed spectra, comparing its recovery of absorber locations and physical parameters against \textsc{viper}.

The paper is organized as follows. In Section \ref{sec: Simulation and Training Dataset}, we describe the simulation setup and construction of the training dataset. Section \ref{sec: Modeling} presents the model architecture and training procedure. In Section \ref{sec: Results}, we discuss the results of our analysis. Section \ref{sec Discussion and conclusion} provides the discussion and conclusions. Finally, in Section \ref{sec: summary}, we summarize our main findings.

\section{Simulation and Training Dataset}
\label{sec: Simulation and Training Dataset}
We train a machine learning model on synthetic Ly$\alpha$ forest spectra and evaluate its performance on high-resolution UVES observations. The training data are generated from cosmological simulations, while the final testing is carried out on real spectra. The details of both the training and observational datasets are described below.

\subsection{Simulation}

The training data used in this work are generated from the TNG100 simulation of the IllustrisTNG project \citep{Nelson2019ComAC...6....2N}, a cosmological magneto-hydrodynamical simulation that follows the evolution of dark matter and baryons from high redshift ($z \sim 127$) to the present day within a periodic box of side length $\sim110~\mathrm{Mpc}$. 
The simulation assumes a standard $\Lambda$CDM cosmology and evolves the coupled dynamics of gas, stars, dark matter, and black holes under gravity and hydrodynamics. 
Gravity is computed using a hybrid Tree–Particle-Mesh (TreePM) scheme, while the gas dynamics are solved using the moving-mesh code \textsc{arepo} \citep{Weinberger2020ApJS..248...32W}. 

In \textsc{arepo}, the gas is discretized on an unstructured mesh defined by a Voronoi tessellation of mesh-generating points. 
Each Voronoi cell corresponds to the region of space closer to its generating point than to any other, forming a space-filling set of irregular polyhedral cells. 
The hydrodynamic equations are solved using a second-order finite-volume Godunov scheme with an approximate Riemann solver, which computes fluxes across cell interfaces. 
The mesh-generating points are advected approximately with the local gas velocity, resulting in a quasi-Lagrangian formulation that reduces advection errors and preserves Galilean invariance compared to fixed-grid methods. This moving-mesh approach combines the advantages of Eulerian and Lagrangian techniques: it accurately captures shocks and discontinuities while naturally adapting spatial resolution to the gas distribution. This is particularly important for modelling the low-density intergalactic medium (IGM), where the Ly$\alpha$ forest arises from smoothly varying density and velocity fields over large scales.

The TNG100 simulation contains $1820^{3}$ dark matter particles and an equal number of gas resolution elements, corresponding to a dark matter particle mass of $\sim7.5\times10^{6}\,M_\odot$ and a baryonic mass resolution of $\sim1.4\times10^{6}\,M_\odot$.
The gravitational softening length for collisionless components is $\sim1~\mathrm{kpc}$ (physical at $z \lesssim 1$), while the gas cells have adaptive spatial resolution that reaches scales of a few kiloparsecs in low-density regions relevant for the Ly$\alpha$ forest.
The simulation includes a comprehensive model for galaxy formation physics, incorporating radiative gas cooling, star formation, chemical enrichment, and feedback from both supernovae and active galactic nuclei (AGN).
Radiative cooling is computed for both primordial and metal-enriched gas in the presence of a spatially uniform, redshift-dependent ultraviolet background based on \citet{Faucher2009ApJ...703.1416F}, 
which determines the ionization state of hydrogen and thus the neutral fraction responsible for Ly$\alpha$ absorption.
Star formation and feedback are both implemented through subgrid models. The stellar feedback is performed via kinetic winds that inject energy, momentum, and metals into the surrounding gas. In addition, AGN feedback operates in multiple modes depending on the accretion state of the central black hole, providing both thermal and kinetic energy input that regulates gas cooling and redistribution on galactic and circumgalactic scales \citep[for more details, refer to][]{Pillepich2018MNRAS.473.4077P}.
These physical processes collectively determine the density, temperature, velocity, and ionization structure of the intergalactic medium, which gives rise to the Ly$\alpha$ forest.
We use simulated spectra for training because they enable the generation of large, homogeneous datasets with consistent quality and well-controlled observational characteristics. While labels are still obtained through Voigt-profile fitting, as for real spectra, simulations provide a self-consistent framework free from many observational complications and selection effects. Furthermore, synthetic observations can be tailored to match the properties of current and future spectroscopic instruments, making them particularly well suited for training CNN-based models.

A number of modern cosmological hydrodynamical simulations, including EAGLE \citep{Crain2015MNRAS.450.1937C}, SIMBA \citep{Dave2019MNRAS.486.2827D}, Nyx \citep{Almgren2013ApJ...765...39A}, and Sherwood \citep{Bolton2017MNRAS.464..897B}, provide physically motivated descriptions of the IGM and are widely used for studies of Ly$\alpha$ absorption. We adopt TNG100 not because it uniquely incorporates these physical ingredients, but because it offers a practical combination of simulation volume, resolution, and publicly available data products suitable for constructing a large set of synthetic sightlines. It is also extensively documented and provides an online interface for accessing and querying simulation products, enabling efficient interaction with the data without requiring the download of large simulation snapshots to local systems. Since the present study focuses on generating realistic training data for machine-learning applications rather than precision constraints on IGM thermodynamics, TNG100 provides an appropriate framework for this purpose.

We note that different feedback prescriptions can lead to variations in the thermal properties of the low-density IGM and warm--hot intergalactic medium (WHIM), which may in principle influence Ly$\alpha$ absorption through thermal broadening and pressure smoothing. However, recent studies comparing Illustris, IllustrisTNG, and other simulations have shown that many observable Ly$\alpha$ absorber properties remain largely insensitive to these differences despite substantial variations in feedback strength and WHIM fractions \citep{2024MNRAS.527.4545K,Hu2024MNRAS.52711338H}. Therefore, while TNG may not reproduce every aspect of IGM thermodynamics identically to alternative simulations, these limitations are not expected to significantly affect its use here as a generator of realistic synthetic sightlines for CNN training.
 
In the next section, we describe how these simulation outputs are used to generate mock absorption spectra using the \textsc{trident} package.

\subsection{Mock Spectra Generation Using \textsc{trident}}

Mock Ly$\alpha$ forest spectra are generated from the simulation outputs  using the \textsc{trident} package \citep{Hummels2017ApJ...847...59H}, which produces synthetic absorption-line spectra by post-processing hydrodynamical simulation data. In this work, we focus on the redshift range $2.5 < z_{abs} < 2.7$, where the Ly$\alpha$ forest is clearly observable in the optical band, while still providing a sufficiently large redshift interval for statistical analysis.
\textsc{trident} first constructs one-dimensional sightlines, referred to as \texttt{LightRays}, by sampling gas properties along a specified path through the simulation volume. 
Each \texttt{LightRays} consists of ordered arrays of the gas density ($\rho$), temperature ($T$), neutral hydrogen number density ($n_{\mathrm{HI}}$), line-of-sight velocity ($v_{\mathrm{los}}$), path length ($\mathrm{d}l$), and effective redshift ($z_{\mathrm{eff}}$) for all gas elements intersected by the ray. To generate sightlines spanning cosmological distances, we construct \textit{compound rays} by stitching together segments from multiple simulation snapshots across the redshift interval $2.5 < z_{abs} < 2.7$, ensuring continuous coverage along the line of sight.

For each \texttt{LightRays}, the \textsc{trident} code computes the number density of neutral hydrogen by interpolating pre-computed ionization tables generated with \textsc{cloudy}. These tables assume ionization equilibrium and incorporate the effect of a spatially uniform, redshift-dependent metagalactic ultraviolet background \citep{Haardt2012ApJ...746..125H}, such that the ionization fractions depend on the local gas density, temperature, and redshift.
The resulting neutral hydrogen density field is then integrated along the line of sight to obtain the column density contribution from each gas element.
The absorption spectrum is constructed by computing the optical depth as a function of wavelength. For each gas element intersected by the sightline, \textsc{trident} deposits an absorption feature using a Voigt profile, which arises from the convolution of Doppler (Gaussian) and natural (Lorentzian) broadening mechanisms. The optical depth for each transition is given by
\begin{equation}
\tau(\lambda) = \tau_0 \, V(\lambda),
\end{equation}
where $V(\lambda)$ is the Voigt profile and the central optical depth is
\begin{equation}
\tau_0 = \frac{\pi e^2}{m_e c} \, N \, f \, \lambda_0,
\end{equation}
with $N$ the column density, $f$ the oscillator strength, and $\lambda_0$ the rest-frame wavelength of the transition \citep{Hummels2017ApJ...847...59H}. The wavelength of each absorber is shifted according to both cosmological redshift and the line-of-sight peculiar velocity of the gas.
The total optical depth along the sightline is obtained by summing the contributions from all gas elements, and the transmitted flux is then computed as
\begin{equation}
F(\lambda) = \exp[-\tau(\lambda)].
\label{ew: tauflux}
\end{equation}

To match the resolution of high-resolution spectrographs such as UVES, we adopt an instrumental broadening corresponding to a velocity full width at half maximum of $\mathrm{FWHM}_v \approx 6~\mathrm{km\,s^{-1}}$, consistent with a resolving power $R \sim 45{,}000$--$55{,}000$. At the mean observed wavelength of the Ly$\alpha$ forest in our redshift range ($z_{abs} \approx 2.6$), this corresponds to a FWHM of 0.0876 \AA. 
%
%
The synthetic spectra are generated on a uniform wavelength grid with spacing $\Delta\lambda = 0.029$~\AA. This choice ensures that each resolution element is sampled by approximately three pixels, exceeding the minimum Nyquist requirement of two samples per resolution element and thereby avoiding undersampling of the instrumental line profile.
%
%

The raw spectra are then convolved with a Gaussian line-spread function (LSF) corresponding to the adopted instrumental resolution. For a Gaussian profile, the standard deviation is related to the full width at half maximum by $\mathrm{FWHM} = 2.355 \sigma$. Since \textsc{trident} performs convolution in wavelength space, the Gaussian kernel width translates into $\sigma_{\mathrm{bins}} \approx 1.29~\text{pixels}$, which is used for the convolution. 
%
%
This method ensures that the applied LSF correctly reproduces the desired velocity resolution on the discretized wavelength grid.

In addition, \textsc{trident} implements a subgrid deposition scheme to account for absorption features narrower than the bin width, preserving total optical depth and flux even when the intrinsic line width is smaller than the spectral resolution \citep{Hummels2017ApJ...847...59H}. Finally, Gaussian noise is added to the synthetic spectra, with signal-to-noise ratios drawn uniformly over the range $\mathrm{S/N}=5-100$. This range was chosen to broadly encompass the quality of the UVES spectra used in this work.
It is important to note that \textsc{trident} does not perform full radiative transfer; instead, it assumes ionization equilibrium and a uniform ultraviolet background, neglecting local radiation sources and non-equilibrium effects. While these assumptions can introduce uncertainties in low-density regions of the IGM, they provide a computationally efficient and widely adopted framework for generating realistic absorption spectra.

\begin{figure*}
    \centering
    \includegraphics[width=\linewidth]{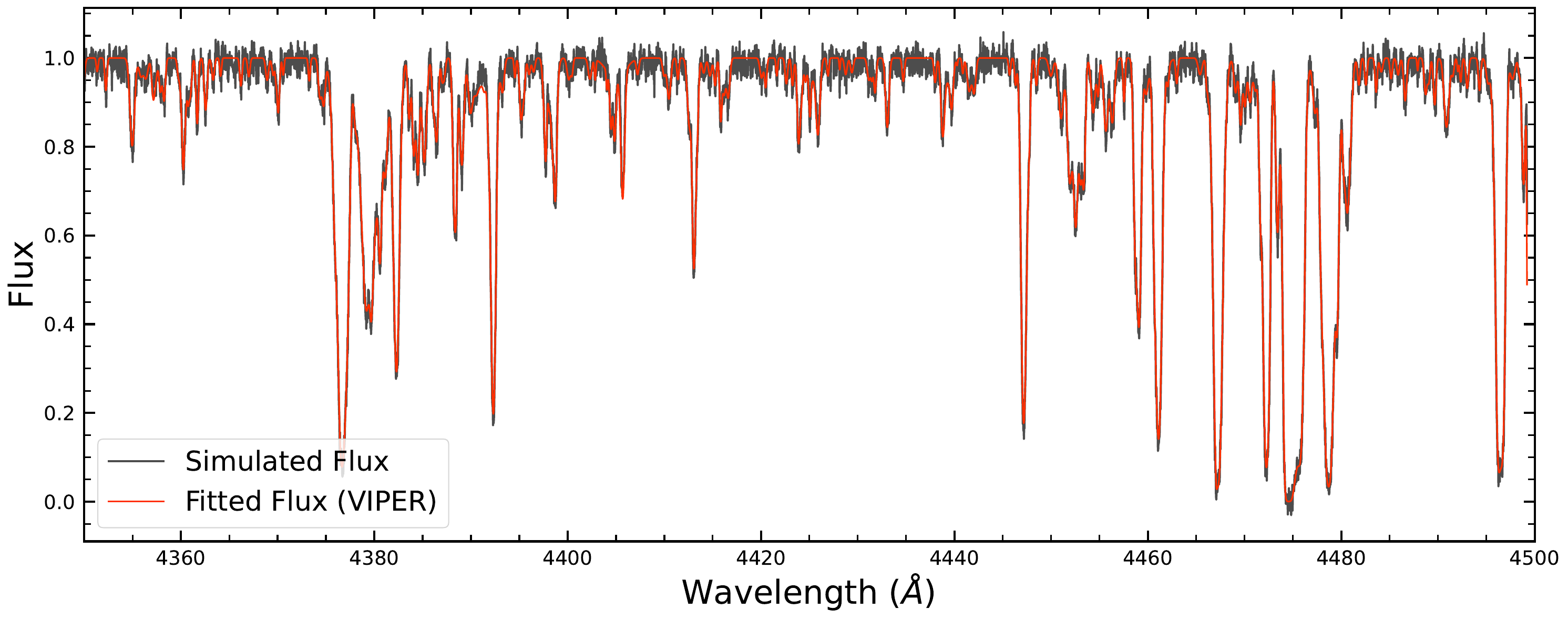}
    \caption{Example of a simulated Ly$\alpha$ forest spectrum generated from the TNG simulation using \textsc{trident}, overlaid with the corresponding Voigt profile fit obtained using \textsc{viper}.}
    \label{fig:simandfitted}
\end{figure*}

The resulting dataset consists of high-resolution synthetic Ly$\alpha$ forest spectra that directly encode the underlying physical properties of the simulated IGM and serve as the training set for our CNN model. An example spectrum is shown in Figure~\ref{fig:simandfitted}, where the simulated flux (black) exhibits a dense series of absorption features arising from intervening neutral hydrogen along the line of sight. The overlaid Voigt profile fit obtained using \textsc{viper} (red) closely reproduces both the depth and width of individual absorption lines, including blended systems, demonstrating the ability of parametric models to recover the underlying absorption structure from the synthetic data.

\subsection{Labeling the Training Data}

\subsubsection{\textsc{viper}: Voigt profile Parameter Estimation Routine}

To obtain physically motivated labels for the training data, each synthetic spectrum is analyzed using the \textsc{viper} (Voigt profile Parameter Estimation Routine) code \citep{Gaikwad2017MNRAS.467.3172G}, which performs automated identification and Voigt profile fitting of absorption features.

\textsc{viper} models absorption lines as Voigt profiles, which arise from the convolution of thermal (Gaussian) broadening and natural (Lorentzian) broadening. For a given transition, the optical depth profile is determined by three physical parameters: the absorber redshift $z_{abs}$, the column density $N_{\mathrm{HI}}$, and the Doppler parameter $b_{\mathrm{HI}}$, which characterises the line width due to thermal and turbulent motions.
The \textsc{viper} algorithm first identifies candidate absorption regions using a significance-based detection scheme. Specifically, it scans the normalized spectrum to locate contiguous segments where the flux decrement exceeds a predefined threshold relative to the noise. These regions are identified based on a \textit{candidate significance level} (CSL), which quantifies the statistical significance of absorption features by comparing their equivalent width to the local noise level.

For each candidate region, initial guesses for absorber parameters are obtained by fitting individual Gaussian profiles to the flux minima within the region. The parameters are then refined through a non-linear least-squares optimisation that minimises the chi-square statistic,
\begin{equation}
\chi^2 = \sum_i \frac{\left[F_{\mathrm{obs}}(\lambda_i) - F_{\mathrm{model}}(\lambda_i)\right]^2}{\sigma_i^2},   
\end{equation}
where $F_{\mathrm{obs}}$ and $F_{\mathrm{model}}$ are the observed and model fluxes, respectively, and $\sigma_i$ is the flux uncertainty. The parameters are allowed to vary within physically motivated bounds (e.g., $5 \lesssim b_{\mathrm{HI}} \lesssim 150~\mathrm{km\,s^{-1}}$ and $10 \lesssim \log N_{\mathrm{HI}} \lesssim 16.5$).

The fitting is performed iteratively, starting with a single-component Voigt profile and progressively increasing the number of components to account for blended or complex absorption features. The optimal number of components is determined using the Akaike Information Criterion with correction (AICC),
\begin{equation}
\mathrm{AICC} = \chi^2 + \frac{2pn}{n - p - 1},   
\end{equation}
where $n$ is the number of data points and $p$ is the number of free parameters in the model. This criterion balances goodness-of-fit against model complexity, penalising overfitting by discouraging unnecessary components. A model is accepted as an improvement only if it results in a sufficiently significant reduction in AICC.

After determining the best-fit model, the significance of individual components is re-evaluated using a \textit{Rigorous significance level} (RSL), which measures the statistical robustness of each component within the full multi-component fit. Components that do not satisfy the RSL threshold are discarded, ensuring that the final catalogue contains only statistically significant absorbers. 

This two-stage procedure—initial detection via CSL, model selection via AICC, and final validation via RSL—allows \textsc{viper} to robustly identify and decompose absorption systems into multiple physically distinct components, each characterised by $(z_{abs}, N_{\mathrm{HI}}, b_{\mathrm{HI}})$.

The final output of \textsc{viper} is a catalogue of absorption components for each spectrum, providing the redshift, column density, and Doppler parameter of all statistically significant Ly$\alpha$ absorbers. Figure \ref{fig:simandfitted} shows an example simulated spectrum fitted with \textsc{viper}. These fitted parameters form the basis for constructing the supervised labels used to train the CNN, as described in the next subsection.

\subsubsection{Construction of Training Labels}
The absorber catalogue generated by \textsc{viper} is used to construct labels for training the CNN, following the general multi-task, sliding-window framework of \citet{Cheng2022MNRAS.517..755C}, with several modifications introduced in this work. We do not directly derive absorber labels from the simulation outputs because the synthetic spectra generated by \textsc{trident} arise from the combined absorption of many gas elements along a line of sight. Consequently, a single absorption feature may result from the superposition of multiple nearby structures, making it difficult to uniquely associate the observed feature with a specific set of simulation elements and assign absorber parameters such as $N_{\rm HI}$, $b$, and redshift. Since our objective is to train the CNN to recover the properties of observable absorption systems, we instead fit the synthetic spectra using \textsc{viper}. This yields labels that correspond directly to the fitted absorption features and are defined in the same manner as those obtained from observational spectra.

Each synthetic spectrum is sampled on a uniform wavelength grid with spacing $\Delta\lambda = 0.029$~\AA,
and labels are assigned at the pixel level. For every pixel, four target quantities are defined: (i) a binary Ly$\alpha$ identification flag (\textit{LyID}), (ii) the neutral hydrogen column density $\log N_{\mathrm{HI}}$, (iii) the Doppler parameter $\log b_{\mathrm{HI}}$, and (iv) a centroid-offset parameter $z_{\mathrm{loc}}$, measured in pixels, which encodes the pixel’s relative position with respect to the absorber center.

The absorber centers are obtained from the fitted redshifts provided by \textsc{viper} and converted to observed-frame wavelengths. These wavelengths are then associated with the discrete spectral grid by identifying the nearest pixel index to each fitted wavelength, ensuring a one-to-one correspondence between each absorber and a central pixel in the spectrum \footnote{While the central pixel is defined via the nearest-grid assignment, the centroid-offset parameter $z_{\mathrm{loc}}$ is treated as a floating-point quantity (in units of pixels) to account for the fact that the true absorption-line center does not, in general, coincide exactly with the center of a pixel.}.

Instead of assigning labels across the full absorption profile, we adopt a localized labeling strategy in which labels are assigned only to the central pixel and its two immediate neighbours (one on either side). Thus, each absorber contributes labels to exactly three pixels. For these pixels, the \textit{LyID} 
flag is set to 1, while $\log N_{\mathrm{HI}}$ and $\log b_{\mathrm{HI}}$ are assigned the values obtained from the \textsc{viper} fit. The centroid-offset label $z_{\mathrm{loc}}$ is defined as the signed pixel distance from the absorber center, taking values $z_{\mathrm{loc}} = 0$ at the central pixel, $z_{\mathrm{loc}} = -1$ for pixels on the blueward side, and $z_{\mathrm{loc}} = +1$ for pixels on the redward side.

All remaining pixels in the spectrum are treated as non-absorbers, with \textit{LyID} = 0 and 
the corresponding physical labels are set to zero. This sparse labeling ensures that only pixels directly associated with absorption features carry physical information, while the majority of the spectrum represents the background.

\begin{figure*}
    \centering
    \includegraphics[width=\linewidth]{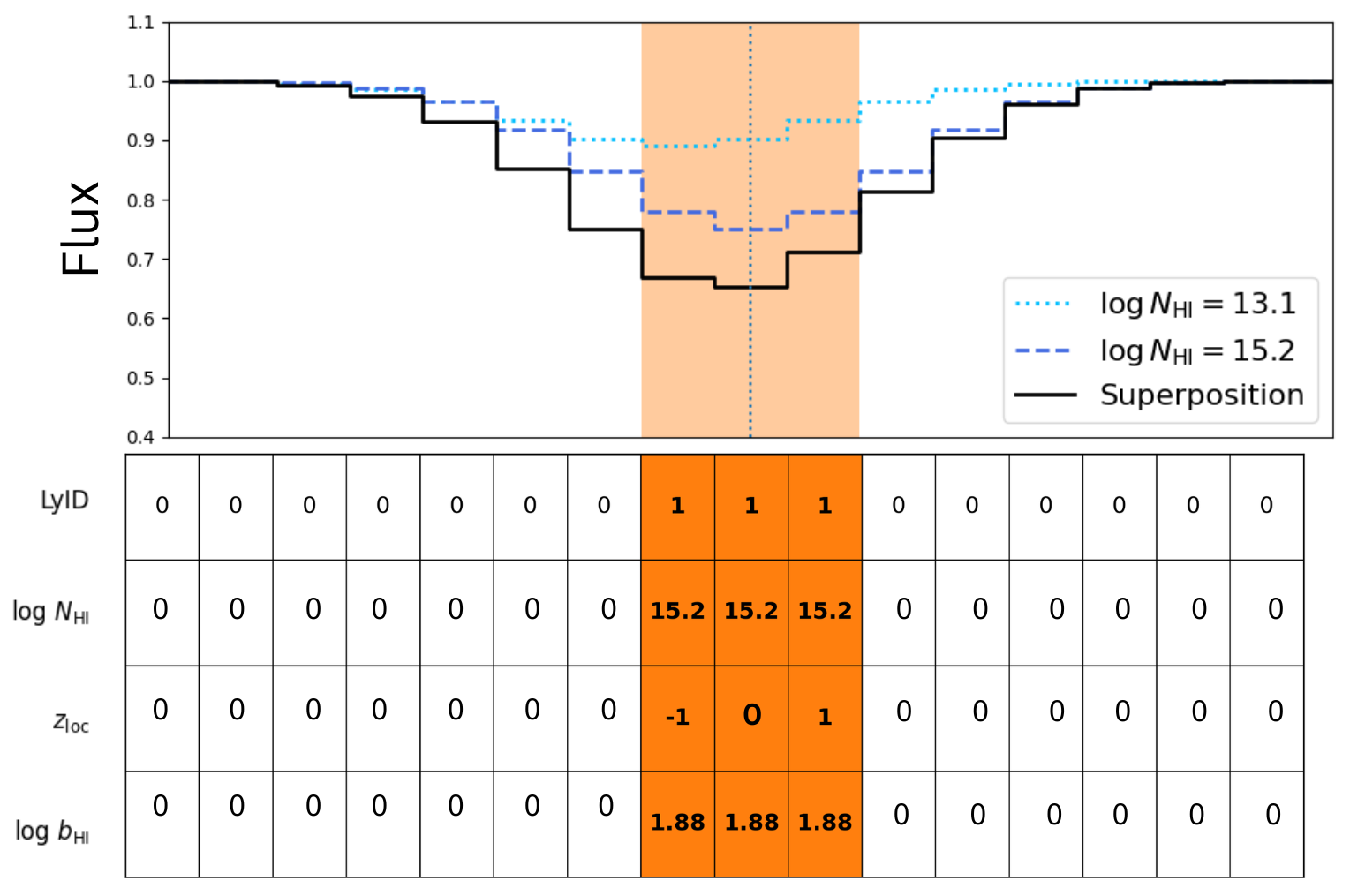}
    \caption{
Illustration of the sliding-window labeling scheme used in this work. A fixed-size window is scanned across the spectrum with a step of one pixel. Labels are assigned only to the central three pixels around an absorption feature, where Ly$\alpha$ absorption is present. Pixels within this region are assigned non-zero values for \textit{LyID}, $\log N_{\mathrm{HI}}$, and $\log b_{\mathrm{HI}}$, while the centroid-offset parameter $z_{\mathrm{loc}}$ takes values $z_{\mathrm{loc}} = 0$ at the central pixel and non-zero values for the adjacent pixels. All other pixels are assigned zero labels. When multiple absorbers are present within the same window, labels are assigned in order of increasing column density, such that the absorber with the highest $\log N_{\mathrm{HI}}$ is associated with the central labeled pixel.
}
\label{fig:labeling_scematic}
\end{figure*}

In cases where multiple absorbers overlap within the same pixel range (i.e., blended systems), labels are assigned in order of increasing column density such that stronger absorbers overwrite weaker ones. This choice ensures that the dominant physical component is retained in regions of overlap and avoids assigning conflicting labels to the same pixel. A schematic diagram of our labeling procedure is shown in Figure \ref{fig:labeling_scematic}.

The final dataset is organised as a set of arrays containing the wavelength grid, flux values, and four corresponding label arrays—Ly$\alpha$ identification flag (\textit{LyID}), $\log N_{\mathrm{HI}}$, $\log b_{\mathrm{HI}}$, and the centroid-offset parameter $z_{\mathrm{loc}}$—for all spectra. For each spectrum, the flux array has length $N_{\mathrm{pix}}$, while each label is stored as an array of the same size. These arrays are subsequently used to construct input-output pairs for the CNN using a sliding-window approach, where a fixed-width window is moved across the spectrum and the labels associated with the central pixel of each window serve as the target outputs.

\subsection{Observational Data}

To evaluate the performance of our model on real data, we use high-resolution quasar spectra from the UVES Spectral Quasar Absorption Database (SQUAD) Data Release 1 \citep{Murphy2019MNRAS.482.3458M}. This dataset consists of 467 quasar spectra observed with the Ultraviolet and Visual Echelle Spectrograph (UVES) on the Very Large Telescope (VLT), providing the spectral resolution and signal-to-noise required for detailed Ly$\alpha$ forest studies.
From this parent sample, we select all quasars with redshift $2.5 < z_{\rm QSO} < 3$, yielding 62 spectra. For each quasar, we extract the Ly$\alpha$ forest region between the Ly$\alpha$ and Ly$\beta$ emission lines in the quasar rest frame. Regions affected by damped Ly$\alpha$ absorbers (DLAs) and the quasar proximity zone are excluded to minimize contamination from strong absorption systems and local ionizing radiation.

The UVES spectra have a resolving power corresponding to a velocity FWHM of approximately $6~\mathrm{km~s^{-1}}$, consistent with the instrumental resolution adopted for the synthetic training spectra. This ensures that the simulated and observational datasets are closely matched in spectral resolution.
All spectra are continuum-normalized prior to analysis. To ensure consistency with the training data, the spectra are resampled onto the same uniform wavelength grid used for the synthetic spectra, with spacing $\Delta\lambda = 0.029$~\AA. This preprocessing step allows the observational data to be directly used as input to the trained CNN without introducing systematic differences in sampling.
The resulting set of continuum-normalized Ly$\alpha$ forest segments extracted from the UVES spectra provides a realistic testbed for assessing the performance of the model on observed data and evaluating its ability to generalize beyond the simulated training set.

\section{Modeling}
\label{sec: Modeling}

Our model follows the general multi-task, sliding-window convolutional neural network (CNN) framework introduced by \citet{Cheng2022MNRAS.517..755C}, with modifications in the labeling strategy described in Section~2.3. The network is designed to simultaneously identify Ly$\alpha$ absorption features and predict their physical parameters from one-dimensional spectra.

\subsection{CNN Architecture}

\begin{figure*}
    \centering
    \includegraphics[width=\textwidth]{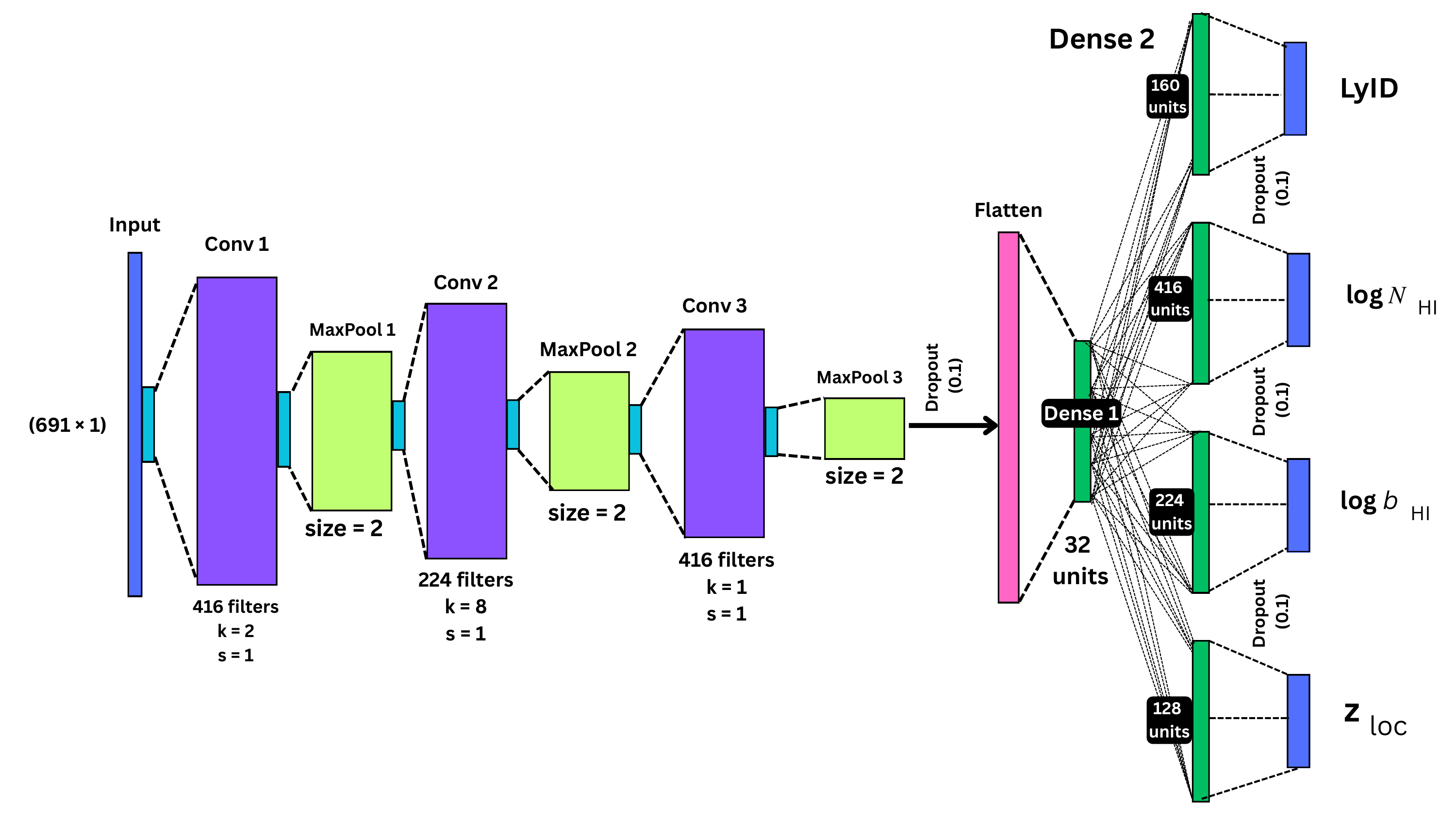}
\caption{Schematic of the one-dimensional convolutional neural network (CNN) used to identify and characterize Ly$\alpha$ absorption features. The network takes a normalized flux window of size $691 \times 1$ as input and processes it through three convolutional layers with 416, 224, and 416 filters, respectively. All convolutional layers use a stride of $s=1$ and are each followed by a max-pooling layer of size 2. The resulting feature maps are flattened and passed through a fully connected layer with 32 units. The network then branches into four task-specific output heads, corresponding to Ly$\alpha$ identification (\textit{LyID}), $\log N_{\rm HI}$, $\log b_{\rm HI}$, and the absorber redshift location ($z_{\rm loc}$). The second dense layers in the four branches contain 160, 416, 224, and 128 units, respectively. Dropout ($p=0.1$) is applied after the convolutional backbone and within each output branch to reduce overfitting. 
}
    \label{fig:cnn_architecture}
\end{figure*}

Figure~\ref{fig:cnn_architecture} illustrates the CNN architecture adopted in this work, which closely follows the framework of \citet{Cheng2022MNRAS.517..755C}. The network takes as input a one-dimensional flux window of size 691 pixels. The network is composed of a sequence of one-dimensional convolutional layers designed to extract hierarchical features from the input flux window.

The convolutional backbone consists of three 1D convolutional layers (Conv1, Conv2, Conv3) 
with filter sizes $(416, 224, 416)$ and kernel widths $(2, 8, 1)$, respectively (Table~\ref{table:optimized_parameters}). Each convolutional layer is followed by a max-pooling layer with a kernel size of 2, which progressively reduces the dimensionality of the feature maps while retaining the most salient features of the absorption profiles. A dropout layer with a rate of 0.1 
is applied after the final pooling layer to mitigate overfitting.

All convolutional and dense layers use the rectified linear unit (ReLU) activation function,
\begin{equation}
    f(x) = \max(0, x),
\end{equation}

which introduces non-linearity while maintaining efficient gradient propagation \citep{Agarap2018arXiv180308375A}.

The output of the convolutional backbone is flattened and passed to a fully connected layer of size 32 (Dense1), which learns a compact representation of the spectral features. This shared representation is then fed into four separate 
fully connected layer
branches corresponding to the four prediction tasks:(\textit{LyID}), $\log N_{\mathrm{HI}}$, $\log b_{\mathrm{HI}}$, $z_{\mathrm{loc}}$.
Each branch contains an additional dense layer followed by dropout, allowing task-specific feature refinement while reducing overfitting.

The activation functions applied to the final output layers of each task-specific branch are chosen based on the nature of the prediction task. For the \textit{LyID} classification task, we use a sigmoid activation function,
\begin{equation}
   f(x) = \frac{1}{1 + e^{-x}}, 
\end{equation}

which outputs a probability in the range $[0,1]$. For the regression tasks, we adopt linear activation functions,
\begin{equation}
    f(x) = x,
\end{equation}
allowing the network to predict continuous values without restriction. These activation functions are applied only at the final layer of each fully connected branch, and not to the intermediate layers.

Several key hyperparameters of the network, including the window size, number of filters, kernel sizes, dense-layer widths, and dropout rate, are optimised using Bayesian optimisation \citep{Snoek2012arXiv1206.2944S} over a predefined search space. The final adopted configuration is summarised in Table~\ref{table:optimized_parameters}.

The model is trained using the Adam optimiser \citep{Kingma2014arXiv1412.6980K} with a learning rate of $10^{-4}$. Training is performed for 50 epochs without applying an explicit early stopping criterion. Instead, we monitor the validation loss during training and retain the model corresponding to the minimum validation loss as the primary trained network. For completeness, we also save the final model obtained at the end of training.

\begin{table}
    \centering
    \caption{Optimized CNN architecture and hyperparameter values adopted in this work.}
    \begin{tabular}{|l|l|}
        \hline
        \textbf{Parameter} & \textbf{Value} \\
        \hline
        Input window size (ws) & 691 \\
        Dropout rate & 0.1 \\
        Max-pooling size & 2 \\
        \hline
        Conv1 filters & 416 \\
        Conv1 kernel size & 2 \\
        Conv2 filters & 224 \\
        Conv2 kernel size & 8 \\
        Conv3 filters & 416 \\
        Conv3 kernel size & 1 \\
        \hline
        Dense1 (shared) & 32 \\
        Dense (\textit{LyID}) & 160 \\
        Dense ($\log N_{\mathrm{HI}}$) & 416 \\
        Dense ($\log b_{\mathrm{HI}}$) & 224 \\
        Dense ($z_{\mathrm{loc}}$) & 128 \\
        \hline
    \end{tabular}
    \label{table:optimized_parameters}
\end{table}

\subsection{Loss Functions}

Our multi-task learning model produces four outputs for each input window: Ly$\alpha$ identification (\textit{LyID}), $\log N_{\mathrm{HI}}$, $z_{\mathrm{loc}}$, and $\log b_{\mathrm{HI}}$. During training, the total loss per epoch is computed as the average over all sliding-window samples.

For the \textit{LyID} classification task, we use the binary cross-entropy loss,
\begin{equation}
    \mathcal{L}_{\mathrm{ID}} = - \frac{1}{N} \sum_{i=1}^{N} \left[ y_{c,i} \log(p_{c,i}) + (1 - y_{c,i}) \log(1 - p_{c,i}) \right],
\end{equation}

where $N$ is the total number of training windows per epoch, $y_{c,i}$ is the true classification label (\textit{LyID} = 1 for an absorber and 0 otherwise), and $p_{c,i}$ is the predicted probability.

For the regression tasks ($\log N_{\mathrm{HI}}$, $z_{\mathrm{loc}}$, and $\log b_{\mathrm{HI}}$), we adopt a masked mean squared error (MSE) loss,
\begin{equation}
    \mathcal{L}_{j} = \frac{1}{N'} \sum_{i=1}^{N} y_{c,i} \left( y_{j,i} - \hat{y}_{j,i} \right)^2,
\end{equation}
where $j \in \{N_{\mathrm{HI}}, z_{loc}, b_{\mathrm{HI}}\}$, $y_{j,i}$ and $\hat{y}_{j,i}$ are the true and predicted values, respectively, and $N'$ is the number of pixels for which $y_{c,i} = 1$. The factor $y_{c,i}$ acts as a mask, ensuring that regression losses are computed only for absorber pixels, while non-absorber pixels do not contribute to the loss.

The total loss function is defined as the sum of the classification and regression losses,
\begin{equation}
   \mathcal{L}_{\mathrm{total}} = \mathcal{L}_{\mathrm{ID}} + \mathcal{L}_{N_{HI}} + \mathcal{L}_{z} + \mathcal{L}_{b_{HI}}. 
\end{equation}


In this work, equal weights are assigned to all loss components. However, due to intrinsic differences in scale, the individual loss terms do not contribute equally to the total loss. From the training curves, we find that the classification loss ($\mathcal{L}_{\mathrm{ID}}$) and the $b_{\mathrm{HI}}$ regression loss are the dominant contributors, while the $z_{\mathrm{loc}}$ loss remains comparatively small.

It is therefore important that the different loss components are brought to comparable scales to avoid biasing the optimisation toward specific tasks. In preliminary experiments, predicting the Doppler parameter $b_{\mathrm{HI}}$ in linear space (with typical values in the range $15$–$75~\mathrm{km\,s^{-1}}$) led to disproportionately large errors, causing the $b_{\mathrm{HI}}$ term to strongly dominate the total loss and degrade the performance of the other tasks. To mitigate this, we instead predict $\log b_{\mathrm{HI}}$, which reduces the dynamic range and leads to a more balanced contribution across tasks.

Although some variation in scale remains, this transformation significantly stabilises the training and prevents any single regression component from overwhelming the optimisation.

\subsection{Evaluation Metrics}

To evaluate the performance of the CNN, we adopt separate metrics for the classification and regression tasks, following \citet{Cheng2022MNRAS.517..755C}.

For the classification of Ly$\alpha$ absorbers, we use \textit{recall}, \textit{precision}, and the \textit{F1 score}, defined as
\begin{equation}
   \mathrm{recall} = \frac{\mathrm{TP}}{\mathrm{TP} + \mathrm{FN}}, \quad
   \mathrm{precision} = \frac{\mathrm{TP}}{\mathrm{TP} + \mathrm{FP}}, 
\end{equation}

and

\begin{equation}
   F_1 = 2 \times\frac{\mathrm{precision}\times\mathrm{recall}}{\mathrm{precision}+\mathrm{recall}}. 
\end{equation}

Here TP (true positives) denotes correctly identified absorbers, FP (false positives) denotes spurious detections, and FN (false negatives) represents true (\textsc{viper}) absorbers missed by the CNN. Recall measures the completeness of detection, while precision quantifies the reliability of the identified systems. The F1 score combines precision and recall into a single metric through their harmonic mean and is particularly useful when assessing the balance between completeness and reliability. A high F1 score indicates that the model simultaneously achieves high recall and high precision.

For the regression tasks, namely the prediction of $\log N_{\mathrm{HI}}$, $\log b_{\mathrm{HI}}$, and absorber location, we use two complementary metrics: the root mean square error (RMSE) and the mean absolute error (MAE). The RMSE is defined as
\begin{equation}
   \mathrm{RMSE} = \sqrt{\frac{1}{N_{\mathrm{abs}}} \sum_{k=1}^{N_{\mathrm{abs}}} (y_k - \hat{y}_k)^2}, 
\end{equation}

where $N_{\mathrm{abs}}$ is the number of matched absorbers, and $y_k$ and $\hat{y}_k$ are the true and predicted values, respectively.

The MAE is defined as
\begin{equation}
    \mathrm{MAE} = \frac{1}{N_{\mathrm{abs}}} \sum_{k=1}^{N_{\mathrm{abs}}} |y_k - \hat{y}_k|.
\end{equation}

The RMSE is more sensitive to outliers due to the quadratic dependence on residuals, while the MAE provides a more robust estimate of the typical prediction error. In this work, we report both metrics, but primarily rely on the MAE when interpreting model performance, as it is less affected by occasional large deviations arising from blended or complex absorption systems.

\subsection{Prediction Uncertainty Estimation}

To estimate uncertainties in the CNN predictions, we employ a Monte Carlo (MC) dropout approach following the Bayesian interpretation of dropout proposed by \citet{Gal2015arXiv150602142G}. During inference, dropout layers are kept active by evaluating the network in training mode, allowing different subsets of neurons to be randomly deactivated during each forward pass. This produces an ensemble of stochastic predictions for the same input spectrum.

For each test spectrum, we perform $N_{\rm MC}$ independent forward passes through the trained CNN. We compared uncertainty estimates obtained using smaller numbers of realizations and found broadly consistent results between $N_{\rm MC}=10$ and $N_{\rm MC}=50$. We therefore adopted $N_{\rm MC}=50$ for all subsequent analyses. For a given predicted quantity $y$ (e.g., $\log N_{\mathrm{HI}}$ or $\log b_{\mathrm{HI}}$), the final prediction is taken as the mean over all realizations,

\begin{equation}
\mu_y=\frac{1}{N_{\rm MC}}
\sum_{i=1}^{N_{\rm MC}} y_i,   
\end{equation}

while the associated predictive uncertainty is estimated from the standard deviation,

\begin{equation}
\sigma_y=
\sqrt{
\frac{1}{N_{\rm MC}}
\sum_{i=1}^{N_{\rm MC}}
(y_i-\mu_y)^2
}.    
\end{equation}

This procedure provides an estimate of the model's epistemic uncertainty, reflecting uncertainty arising from imperfect knowledge of the network parameters rather than measurement noise. Regions where the CNN produces larger predictive scatter correspond to inputs for which the model is less confident. These uncertainty estimates are propagated into subsequent comparisons with fitted absorber catalogs.

\section{Results}
\label{sec: Results}

In this section, we present the performance of our convolutional neural network on both simulated and observational quasar spectra. A total of 500 mock spectra were generated from the TNG simulation, of which 400 were used for training, 50 for validation, and 50 for testing.
During inference, predictions are performed in a sliding-window framework. Each spectrum is padded at both ends to ensure that predictions can be obtained for all pixels, including those near the boundaries. A window of size 691 pixels is then moved across the spectrum one pixel at a time, and the CNN produces a prediction for the central pixel of each window. This procedure yields a complete set of pixel-level predictions for the entire spectrum.
To convert these pixel-level outputs into absorber-level detections, we group contiguous pixels with \textit{LyID} values exceeding a chosen threshold. Within each group, the absorber location is defined as the pixel with the minimum $|z_{\rm loc}|$, corresponding to the estimated line centre. The physical parameters ($\log N_{\rm HI}$ and $\log b_{\rm HI}$) are then assigned using the predictions at this selected pixel.

For evaluation, predicted absorbers are matched to systems identified by \textsc{viper} using a velocity-based criterion. A match is assigned if the velocity separation is less than $\Delta v_{\rm thresh} = 10~{\rm km\,s^{-1}}$.
This threshold is slightly larger than the typical UVES velocity resolution of $\sim6~{\rm km\,s^{-1}}$, allowing for small uncertainties in the recovered absorber centroids while minimizing ambiguous matches between neighbouring absorption systems. Each predicted absorber can be matched to at most one \textsc{viper} absorber and vice versa, ensuring a one-to-one correspondence. Matched systems contribute to the true positive (TP) sample and are used to evaluate parameter recovery, while unmatched CNN predictions and unmatched absorbers in the \textsc{viper} reference catalogue are counted as false positives (FP) and false negatives (FN), respectively.

\begin{figure}
\centering
\includegraphics[width=1\linewidth]{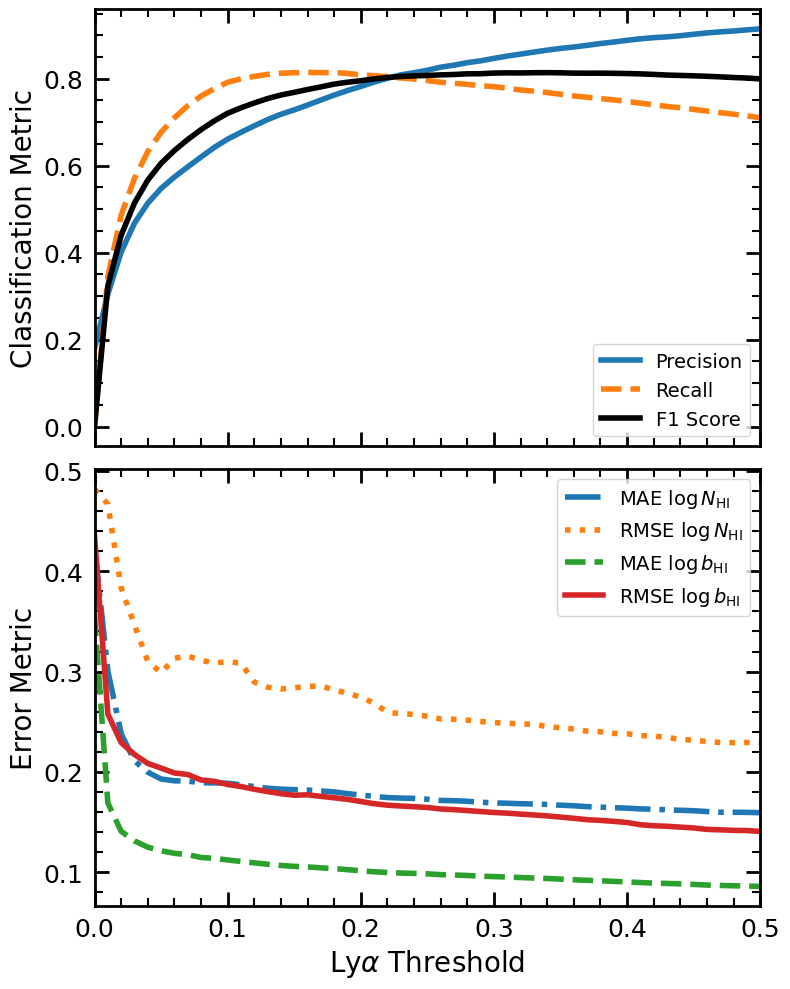}
\caption{
Variation of the evaluation metrics as a function of the \textit{LyID} threshold for simulated spectra.
Top panel: Precision (the fraction of CNN-predicted absorbers matched to the VIPER reference catalogue) and recall (the fraction of VIPER reference absorbers correctly identified by the CNN)}, along with their harmonic mean, the F1 score.
Bottom panel: Mean Absolute Error (MAE) and Root Mean Square Error (RMSE) for the predicted $\log N_{\mathrm{HI}}$ and $\log b_{\mathrm{HI}}$.
\label{fig:evaluation_metrices}
\end{figure}

\begin{figure*}
\centering
\includegraphics[width=0.48\textwidth]{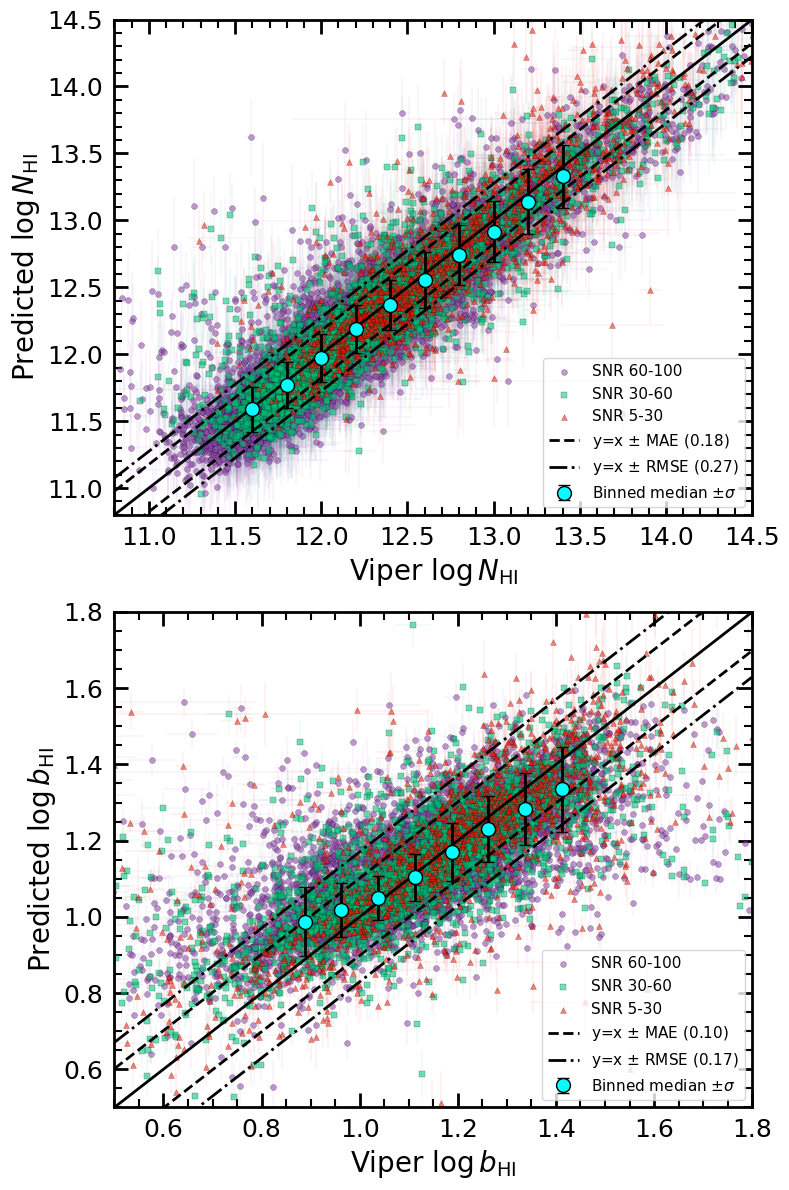}
\hfill
\includegraphics[width=0.48\textwidth]{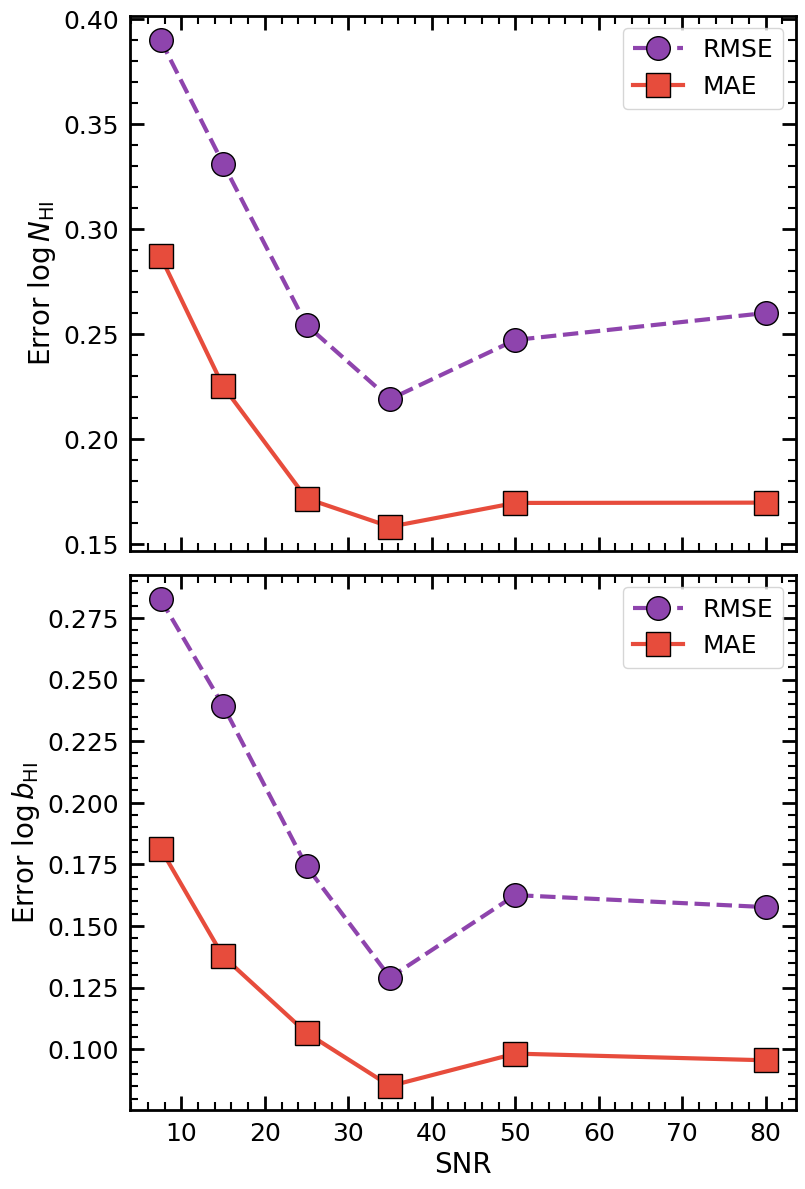}
\caption{Comparison between the predicted and \textsc{viper} reference absorber column densities ($\log N_{\mathrm{HI}}$) and Doppler parameters ($\log b_{\mathrm{HI}}$) for simulated spectra. The left panels show the scatter between the predicted and \textsc{viper} reference values of the column density ($\log N_{\mathrm{HI}}$; top) and the Doppler parameter ($\log b_{\mathrm{HI}}$; bottom). In each panel, the solid black line denotes the one-to-one relation ($y=x$). The dashed blue lines mark the loci $y=x\pm\mathrm{MAE}$ (mean absolute error), while the dashed red lines indicate $y=x\pm\mathrm{RMSE}$ (root mean square error). The data points are shown with uncertainties in both directions, where the x-axis error bars are taken from the \textsc{viper} fitted parameter uncertainties and the y-axis error bars are derived from Monte Carlo dropout estimates of the CNN predictions. The cyan circles represent the median predicted values in bins of the \textsc{viper} parameter, with error bars indicating the corresponding scatter within each bin. The right panels show the dependence of the prediction errors on the signal-to-noise ratio (SNR) of the simulated spectra. The mean absolute error (MAE; red squares) and root mean square error (RMSE; purple circles) are shown for both $\log N_{\mathrm{HI}}$ (top) and $\log b_{\mathrm{HI}}$ (bottom).}

\label{fig:scatterpredvstrue_simulated}
\end{figure*}

Applying the above procedure to the simulated test set, we match the CNN-identified absorbers to those obtained from \textsc{viper}. Figure~\ref{fig:evaluation_metrices} shows the variation of evaluation metrics as a function of the \textit{LyID} threshold. As expected, increasing the threshold improves precision while reducing recall, reflecting the trade-off between completeness and purity. This behaviour arises because higher thresholds suppress low-confidence detections, thereby reducing false positives at the expense of missing weaker absorption systems. 
Figure~\ref{fig:scatterpredvstrue_simulated} left panel compares the predicted absorber properties for the regression tasks ($\log N_{\rm HI}$ and $\log b_{\rm HI}$) with the corresponding matched \textsc{viper} measurements. We adopt a \textit{LyID} threshold of 0.2, which provides a reasonable balance between precision and recall. At this threshold, the CNN successfully recovers the majority of strong absorption systems while maintaining a low contamination rate, and the predicted physical parameters show good agreement with the fitted values. The right panels quantify the dependence of the prediction accuracy on the signal-to-noise ratio (SNR), showing the variation of the MAE and RMSE for both $\log N_{\rm HI}$ and $\log b_{\rm HI}$ across different SNR bins.

\begin{figure}
\centering
\includegraphics[width=0.9\linewidth]{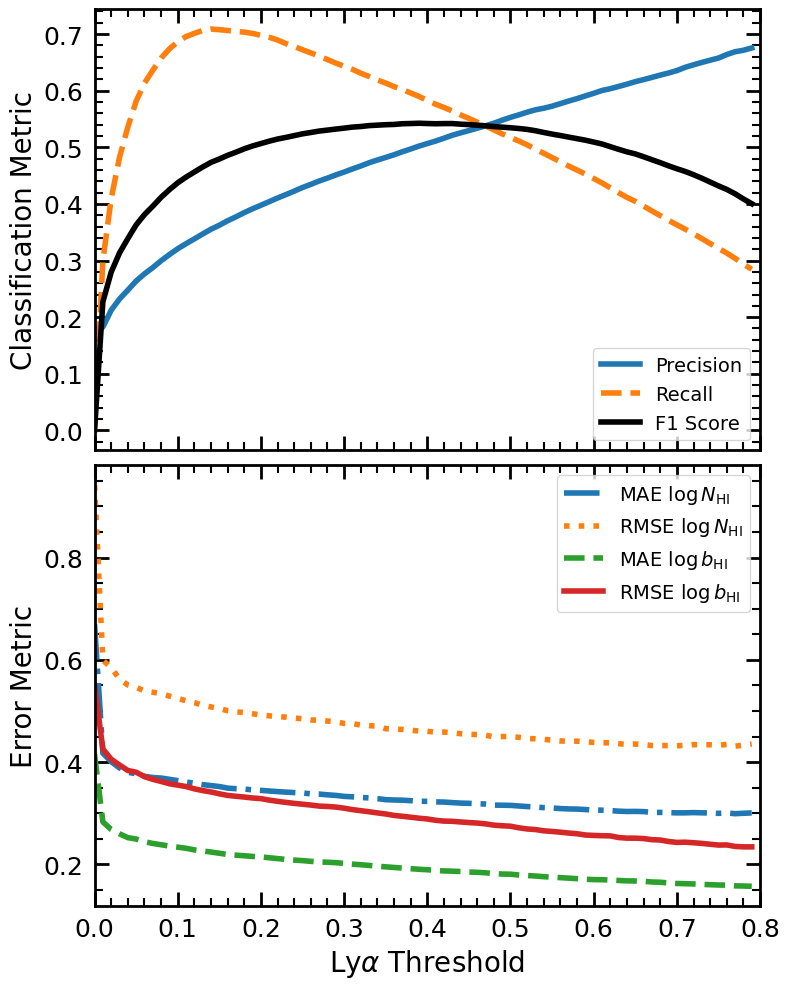}
\caption{
Same as Figure~\ref{fig:evaluation_metrices}, but for the observational UVES spectra.
}
\label{fig:Evaluation_metrices_real}
\end{figure}

\begin{figure*}
\centering
\includegraphics[width=0.48\textwidth]{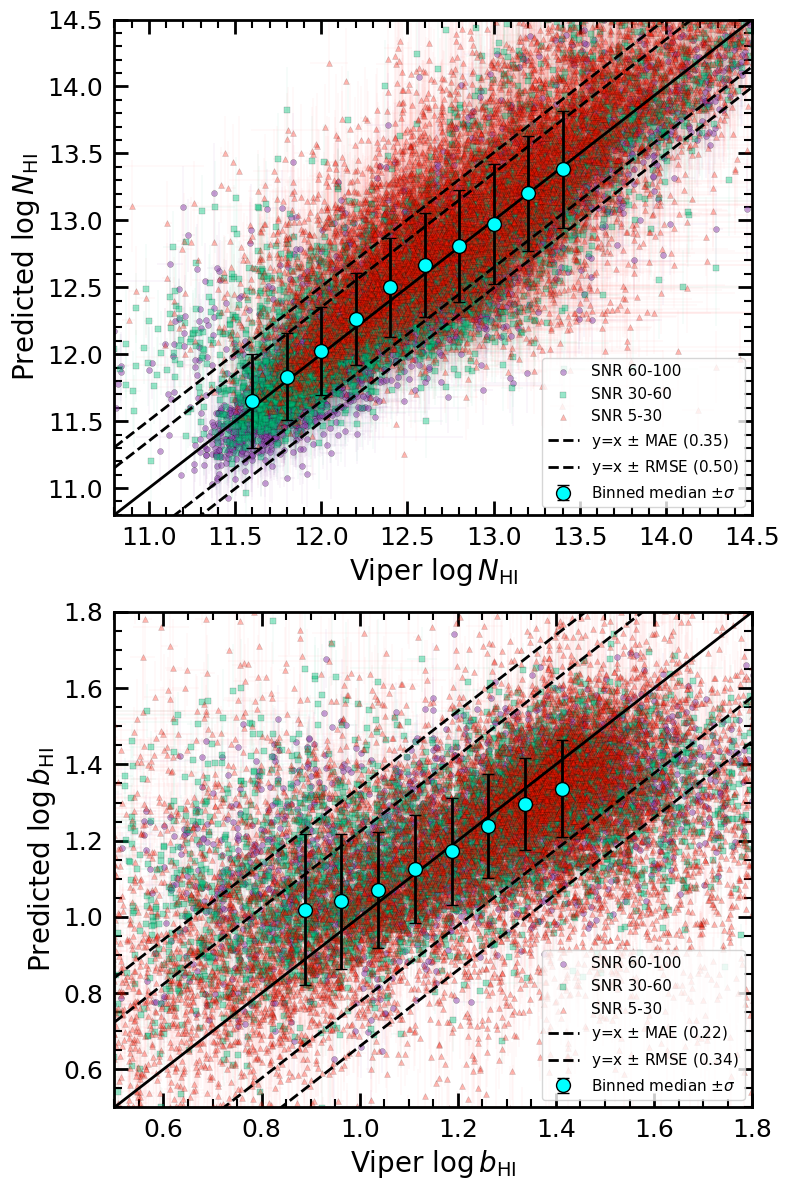}
\hfill
\includegraphics[width=0.48\textwidth]{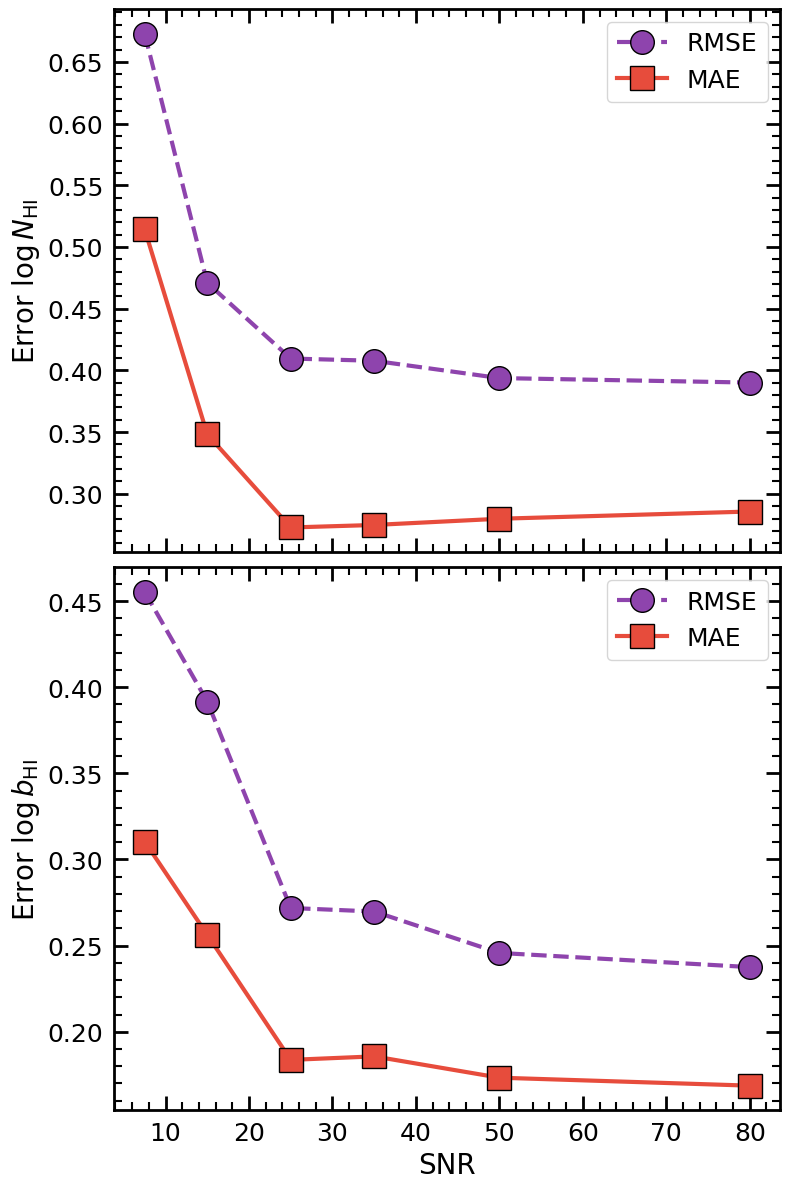}
\caption{Same as Figure~\ref{fig:scatterpredvstrue_simulated}, but for the observational UVES spectra.}
\label{fig:scatterpredvstrue_real}
\end{figure*}

Predictions on the observational spectra were performed using the same sliding-window procedure as for the simulated (TNG) spectra.
Figure~\ref{fig:Evaluation_metrices_real} summarizes the corresponding evaluation metrics—precision, recall, mean absolute error (MAE), and root mean square error (RMSE)—computed across the UVES sample as a function of the \textit{LyID} threshold. As in the simulated case, increasing the threshold leads to the expected trade-off between precision and recall, while the regression metrics quantify the accuracy of the recovered absorber properties. We adopt a \textit{LyID} threshold of 0.2, as determined from the simulated test set, and apply the same threshold to the observational sample. This ensures that the threshold is fixed independently of the observational data and allows the CNN performance on the observational spectra to be evaluated consistently.
 Figure~\ref{fig:scatterpredvstrue_real} left panel further compares the CNN-predicted absorber parameters ($\log N_{\rm HI}$ and $\log b_{\rm HI}$) with the corresponding matched \textsc{viper} measurements. The right panels show the dependence of the prediction errors on the signal-to-noise ratio (SNR) of the spectra, illustrating how the parameter recovery varies with data quality. 

\begin{figure}
\centering
\includegraphics[width=\linewidth]{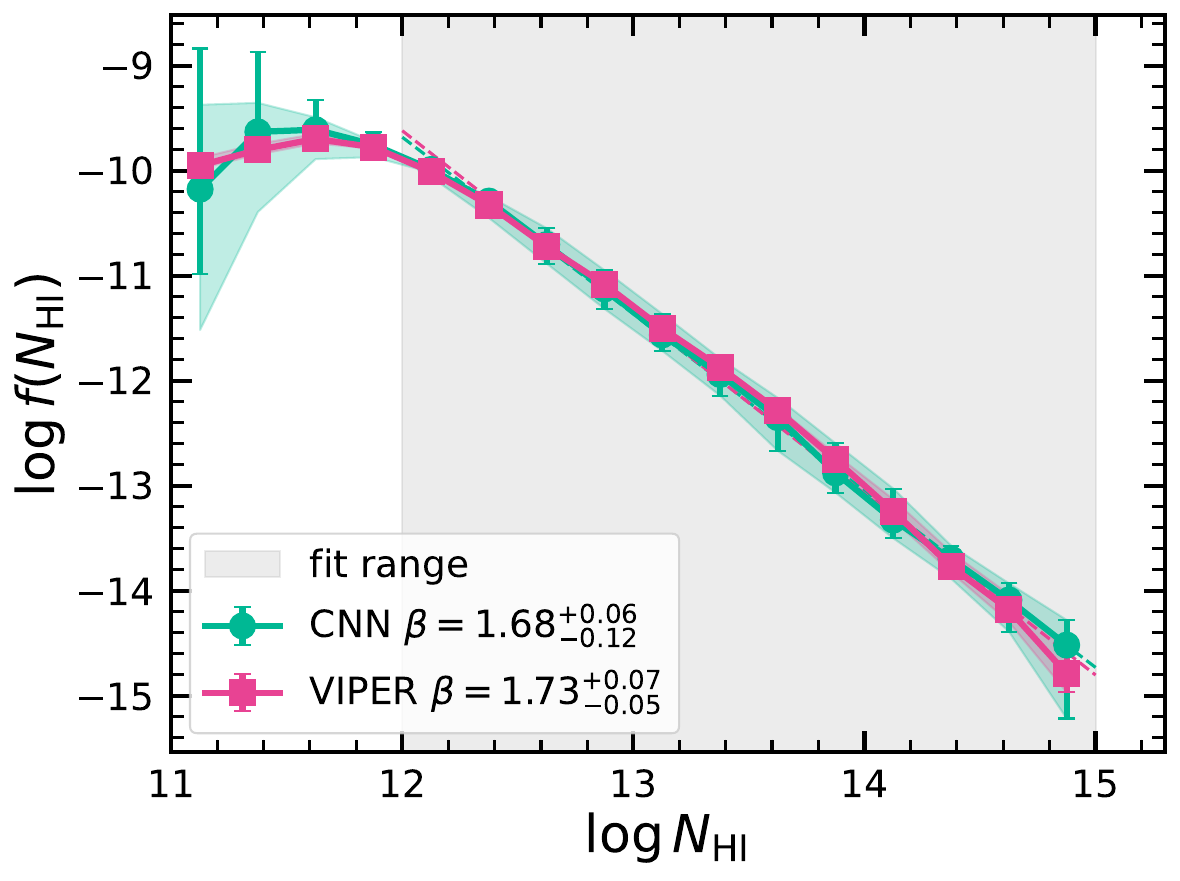}
\includegraphics[width=\linewidth]{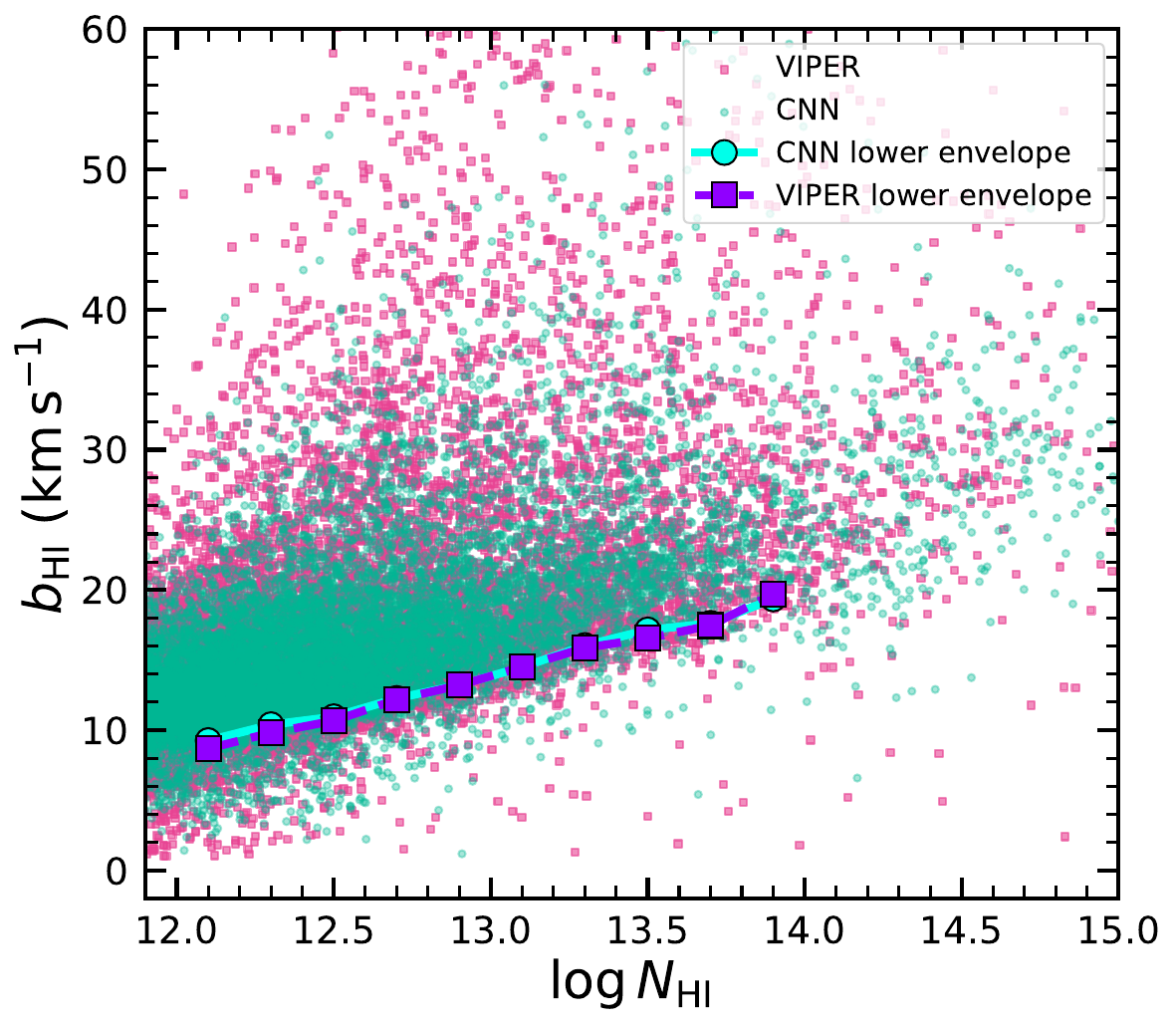}
\caption{Top panel: Comparison of the H\,\textsc{i} column density distribution function (CDDF) derived from the simulated absorber catalogues identified by the CNN (circles) and \textsc{viper} (squares). Bottom panel: Distribution of Doppler parameter ($b_{\rm HI}$), as a function of H\,\textsc{i} column density ($\log N_{\rm HI}$). The lower-envelope relation is computed using the 10th percentile of $b_{\rm HI}$ values within logarithmic $N_{\rm HI}$ bins. Squares denote the \textsc{viper} lower envelope, while circles represent the corresponding CNN-derived lower envelope.
}
\label{fig:CDDF_andbvsn_simulated}
\end{figure}

\begin{figure}
\centering
\includegraphics[width=\linewidth]{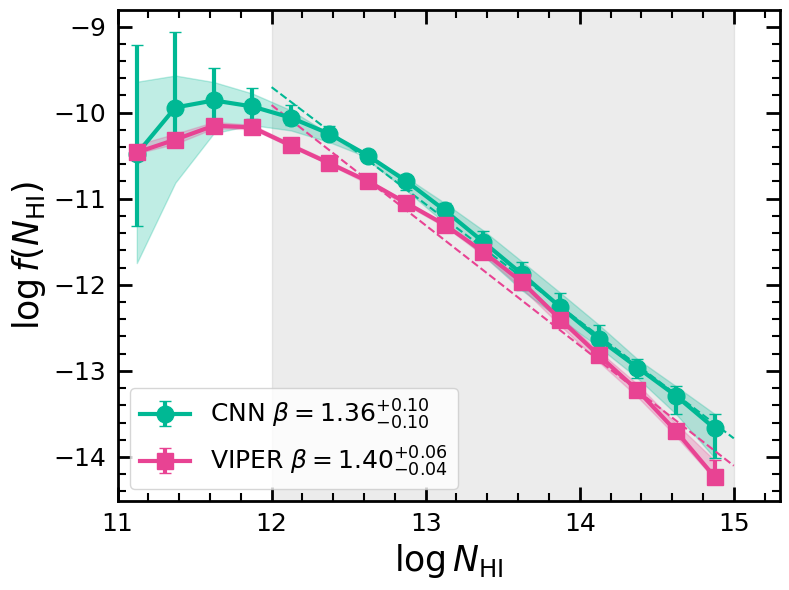}
\includegraphics[width=\linewidth]{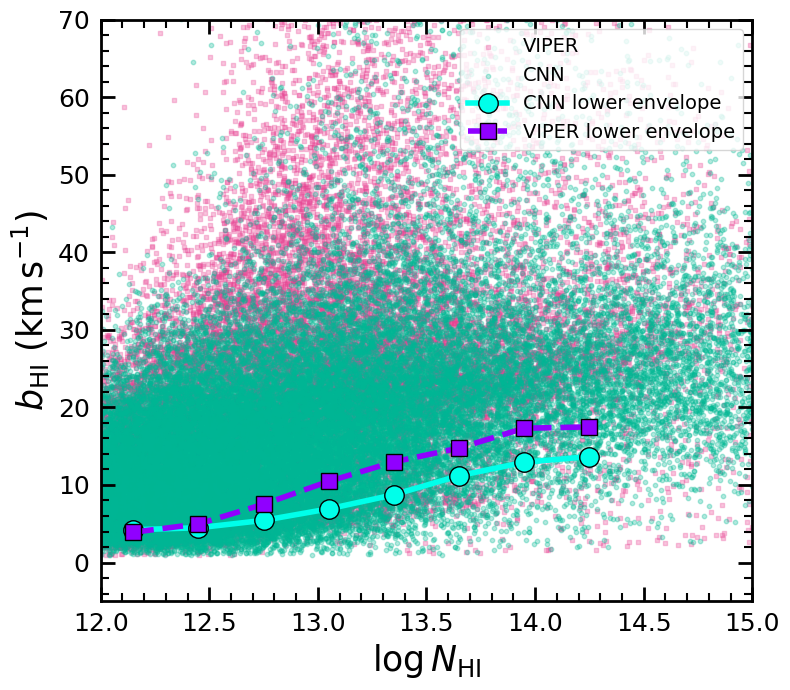}
\caption{
Same as Figure~\ref{fig:CDDF_andbvsn_simulated}, but for the observational UVES spectra.
}
\label{fig:CDDF_andbvsn_real}
\end{figure}

An important test is whether the absorber population identified by the CNN reproduces the statistical properties of the Ly$\alpha$ forest. We therefore examine two commonly used diagnostics: the H~\textsc{i} column density distribution function (CDDF) and the distribution of Doppler parameter as a function of column density ($b_{\rm HI}$--$N_{\rm HI}$). Figure~\ref{fig:CDDF_andbvsn_simulated} (top panel) shows the CDDF derived from CNN-identified absorbers and compares it with the corresponding \textsc{viper} measurements for the simulated spectra. Over the fitted column density range, both distributions are fitted with a power law of the form $f(N_{\rm HI}) \propto N_{\rm HI}^{-\beta}$. The bottom panel shows the $b_{\rm HI}$--$N_{\rm HI}$ distribution obtained from the CNN and \textsc{viper}. In addition to the full distribution, we compute the lower-envelope relation using the 10th percentile of $b_{\rm HI}$ within logarithmic $N_{\rm HI}$ bins and compare the results from the two catalogues.
Figure~\ref{fig:CDDF_andbvsn_real} presents the corresponding comparison for the observational UVES spectra, using the same analysis and fitting procedure as for the simulated data.

The comparison between the MC-dropout uncertainties and the actual prediction residuals provides an important caution for the interpretation of the error bars. For a perfectly calibrated Gaussian uncertainty, approximately 68.3\% of the predictions are expected to lie within $1\sigma$. For $\log N_{\rm HI}$, the MC-dropout uncertainties provide a reasonably good representation of the prediction scatter, with $1\sigma$ coverage fractions of 79.6\% and 59.6\% for the simulated and UVES samples, respectively. In contrast, the uncertainties in $\log b_{\rm HI}$ are substantially underestimated, with corresponding $1\sigma$ coverage fractions of only 28.1\% and 19.5\% for the simulated and UVES samples, respectively. To quantify the required rescaling, we calculated the 68.3rd percentile of the empirical distributions of the absolute normalized residuals, $|y_{\rm CNN}-y_{\rm VIPER}|/\sigma_{\rm MC}$, where $y_{\rm CNN}$ is the CNN-predicted absorber parameter, $y_{\rm VIPER}$ is the corresponding reference value from the VIPER catalogue, and $\sigma_{\rm MC}$ is the standard deviation of the MC-dropout predictions for that parameter. The resulting multiplicative correction factors for $\log N_{\rm HI}$ are 0.74 and 1.24, while those for $\log b_{\rm HI}$ are 3.44 and 4.90, for the simulated and UVES samples, respectively. The stronger underestimation of the $\log b_{\rm HI}$ uncertainties in the UVES sample indicates that additional scatter associated with the simulation-to-observation domain shift is not captured by the MC-dropout uncertainties. We therefore caution that these uncertainties do not represent fully calibrated total uncertainties, particularly for $\log b_{\rm HI}$ in observational spectra. These empirical factors quantify the rescaling required to achieve approximately 68.3\% coverage in each sample and can be used to account for this effect in quantitative applications.

\section{Discussion and Conclusions}
\label{sec Discussion and conclusion}
To assess the overall performance of our model, we compared the CNN predictions against the \textsc{viper} catalog for both simulated and real spectra. On simulated TNG spectra, the CNN reproduces absorber locations and basic Voigt-profile parameters from the \textsc{viper} reference catalogue with good fidelity (see Figure~\ref{fig:evaluation_metrices}). The \textit{LyID} output also provides a tunable confidence measure that allows us to control the trade-off between precision and recall. To estimate uncertainties in the predicted absorber properties, we additionally implemented a Monte Carlo dropout framework during inference. Unlike previous CNN-based approaches for Ly$\alpha$ forest analysis, this enables our model to provide predictive uncertainties together with absorber classifications and physical parameters, allowing the confidence of individual predictions to be quantified.
For the simulated test set, we adopt a representative \textit{LyID} threshold of 0.2, which provides a good balance between precision and recall. The F1 score remains nearly constant at $\sim0.8$ for \textit{LyID} thresholds $\gtrsim0.2$, indicating that the balance between completeness and reliability does not significantly improve at higher thresholds. At this threshold, we obtain a mean absolute error of $\mathrm{MAE} \simeq 0.18$ and a root mean square error of $\mathrm{RMSE} \simeq 0.27$ for $\log N_{\rm HI}$, and $\mathrm{MAE} \simeq 0.10$ and $\mathrm{RMSE} \simeq 0.17$ for $\log b_{\rm HI}$ (see Figure~\ref{fig:scatterpredvstrue_simulated}). The right panels of Figure~\ref{fig:scatterpredvstrue_simulated} show that the prediction errors decrease with increasing signal-to-noise ratio, reaching their lowest values at intermediate SNR ($\sim30$--40) before remaining approximately constant at higher SNR. Although the RMSE exhibits a modest increase at the highest SNR values, the MAE remains nearly unchanged. Visual inspection of the spectra indicates that, at high SNR, \textsc{viper} resolves additional weak Voigt-profile components within individual absorption systems. Since the training labels were constructed by retaining the higher column-density component, the increased number of resolved components occasionally causes the predicted absorber to be matched to a neighbouring Voigt-profile component. These mismatches produce a small number of larger residuals, preferentially increasing the RMSE while leaving the MAE largely unaffected. Overall, these results demonstrate that the CNN performs robustly across a wide range of noise conditions and can reproduce the absorber parameters obtained from the \textsc{viper} reference catalogue.

When applied to real UVES data, the overall performance decreases relative to simulations. Based on the precision--recall trade-off shown in Figure~\ref{fig:Evaluation_metrices_real}, we adopt a \textit{LyID} threshold of 0.2 for the observational sample. At this threshold, the F1 score decreases to approximately $0.5$, indicating a reduced balance between completeness and reliability in real spectra compared to the simulated test set. We obtain $\mathrm{MAE} \simeq 0.35$ and $\mathrm{RMSE} \simeq 0.50$ for $\log N_{\rm HI}$, and $\mathrm{MAE} \simeq 0.22$ and $\mathrm{RMSE} \simeq 0.34$ for $\log b_{\rm HI}$ (see Figure~\ref{fig:scatterpredvstrue_real}). A similar trend is observed in the observational sample, where both the MAE and RMSE decrease with increasing SNR and reach their lowest values at intermediate SNR ($\sim25$--40). Unlike the simulated sample, however, no increase in the RMSE is observed at the highest SNR. Instead, the prediction errors continue to decrease or remain nearly constant. This behaviour likely reflects the fact that, for real spectra, the reduction in observational uncertainties with increasing SNR outweighs the effect of the additional resolved absorption components. Consequently, the factors discussed later in this section dominate the error budget at low SNR and progressively diminish as the SNR increases, leading to the overall downward trend in both the MAE and RMSE.
Overall, the results demonstrate that the CNN maintains consistent performance in reproducing the fitted absorber parameters from \textsc{viper} across the full range of SNR values present in the observational sample.

In addition to comparing the accuracy of individual absorber parameters, it is important to assess whether the CNN preserves the global statistical properties of the Ly$\alpha$ forest. We therefore examined two commonly used diagnostics: the H\,\textsc{i} column density distribution function (CDDF) and the $b_{\rm HI}$--$N_{\rm HI}$ distribution.
For the simulated spectra (see Figure \ref{fig:CDDF_andbvsn_simulated} top panel), the agreement between the CNN and \textsc{viper} catalogues is very close. Over the fitted column density range, both CDDFs are well described by a power-law relation of the form $f(N_{\rm HI})\propto N_{\rm HI}^{-\beta}$. The CNN catalogue yields a slope of $\beta = 1.68^{+0.06}_{-0.12}$, compared to $\beta = 1.73^{+0.07}_{-0.05}$ obtained from the corresponding \textsc{viper} catalogue, corresponding to a difference of only $\Delta\beta \simeq 0.05$. Given the uncertainties, the two measurements are fully consistent.
Next for the $b_{\rm HI}$--$N_{\rm HI}$ distribution, the lower envelope of this relation is of particular physical interest because it traces the minimum line widths present at a given column density. The CNN-derived lower envelope closely follows that recovered by \textsc{viper} across the full column-density range considered, with a root mean square difference of only $0.36\,{\rm km\,s^{-1}}$ between the two relations. The CNN also reproduces the overall distribution of absorbers and the characteristic increase of Doppler parameter with increasing column density.
Taken together, the close agreement in both the CDDF and lower-envelope relation indicates that the CNN preserves the key statistical properties of the Ly$\alpha$ forest as characterized by \textsc{viper}, reproducing the abundance and distribution of its identified absorption systems.

For the observational UVES spectra, the CDDF derived from the CNN catalogue closely follows that obtained from \textsc{viper}, yielding nearly identical power-law slopes of $\beta=1.36^{+0.10}_{-0.10}$ and $\beta=1.40^{+0.06}_{-0.04}$, respectively (see Figure~\ref{fig:CDDF_andbvsn_real}, top panel). The observed slope is also consistent with the value reported by \cite{Kim2013A&A...552A..77K}, who measured $\beta=1.38\pm0.04$ over $\log N_{\rm HI}=12.75$--$14.0$ at $2.4<z<3.2$. For the simulated spectra, we obtain a steeper slope of $\beta\sim1.7$ over the same fitted range, indicating relatively fewer high-column-density absorbers compared with the observations. The CDDF can depend on the adopted ionization and UV-background prescriptions and the treatment of the simulated gas; however, investigating the origin of this difference is beyond the scope of the present work. Our purpose is to use the simulated and observational spectra to assess the CNN against \textsc{viper}, with \textsc{viper} serving as the reference catalogue. Since the same \textsc{viper} analysis is applied to both datasets, the difference between the simulated and observational CDDFs does not affect our primary CNN--\textsc{viper} comparison.

Likewise, the CNN recovers the general structure of the $b_{\rm HI}$--$N_{\rm HI}$ distribution, including the increase of Doppler parameter with increasing column density and the shape of the lower-envelope relation. The lower envelope derived from the CNN predictions remains broadly consistent with that obtained from \textsc{viper}, with a root mean square difference of $3.20\,{\rm km\,s^{-1}}$ between the two relations. Compared to the simulated spectra, however, the agreement is somewhat weaker, particularly at higher column densities ($\log N_{\rm HI}\gtrsim13.5$), where the CNN-derived lower envelope lies systematically below the corresponding \textsc{viper} relation by a few km s$^{-1}$.

Compared to the simulated test set, the CNN exhibits a systematic decrease in performance when applied to real UVES spectra. This degradation is evident in both the recovery of individual absorber parameters, where the MAE, RMSE, and F1 score are poorer than for the simulated sample, and in the statistical diagnostics. Despite this reduction in accuracy, the CNN successfully recovers a substantial fraction of the absorbers identified by \textsc{viper}, reproduces the overall trends of the CDDF and the $b_{\rm HI}$--$N_{\rm HI}$ distributions, and maintains reasonable accuracy in the inferred physical parameters. These results demonstrate that the network generalizes well to observational data, although a measurable performance gap remains.

To investigate the origin of this degradation, we examined whether a domain shift exists between the simulated training data and the real observational spectra. Using the 32-dimensional latent representation learned by the CNN, we projected the training, simulated test, and observational datasets into a common feature space. A principal component analysis shows that the first two principal components account for 98.8\% of the total variance, indicating that the projection accurately captures the latent distribution. The latent representations of the training and simulated test samples are nearly coincident, with a Euclidean centroid distance of only 0.014, demonstrating that the network generalizes well to unseen spectra drawn from the same simulated distribution. In contrast, the observational spectra occupy a largely distinct region of the latent space, with a Euclidean centroid distance of 0.67 from the training distribution. This clear separation provides strong evidence for a significant domain mismatch between the simulated and observational datasets, suggesting that a substantial fraction of the observed performance degradation arises because the network is applied to spectra whose characteristics differ from those encountered during training. Such differences likely reflect observational effects that are not fully represented in the current mock-spectra pipeline, including variations in noise properties across echelle orders, instrumental effects, residual continuum-fitting uncertainties, wavelength-calibration imperfections, and weak metal-line contamination. In addition, our current methodology does not explicitly account for blended absorption systems, which introduces additional scatter in the recovered absorber parameters. Incorporating these effects into future simulations, together with domain-adaptation techniques, should help reduce the remaining performance gap between simulations and observations.


An important advantage of the CNN-based approach is its computational efficiency and its ability to exploit GPU acceleration. For a UVES spectrum containing approximately 25,000 pixels, a single forward pass on a freely available NVIDIA T4 GPU through Google Colab requires approximately 4.0~s. Since our uncertainty-aware analysis uses 50 Monte Carlo dropout realizations, the total CNN inference time is approximately 200~s ($\sim3.3$~min) per spectrum. For comparison, the corresponding \textsc{viper} fit requires 863~s ($\sim14.4$~min) on a 12th Gen Intel Core i5-1235U CPU with 16~GB RAM. The computational cost of \textsc{viper} depends not only on the spectrum length but also on the number and complexity of absorption features that need to be fitted, and can therefore vary substantially between spectra. In contrast, the CNN inference cost is primarily determined by the number of input windows and therefore scales only with spectrum length. Thus, even when accounting for the full 50-realization uncertainty estimation, the CNN processes this spectrum approximately 4.3 times faster than \textsc{viper} using a readily accessible GPU. The CNN naturally exploits GPU parallelism by processing the large number of input windows simultaneously, making it well suited for high-throughput analysis of large spectroscopic samples. \textsc{Viper}, in contrast, is a CPU-based fitting procedure for which the processing of multiple spectra can be parallelized across CPU cores, although the achievable throughput is ultimately limited by the available CPU resources. The CNN throughput can likewise be increased by processing multiple spectra in batches on the GPU, subject to available GPU memory, while more powerful GPUs can provide further acceleration. For reference, on an NVIDIA A100 GPU, the same 50 forward passes require only $\sim40$~s, demonstrating the potential for further acceleration with higher-end GPU hardware.

It is worth noting that, although the absorber labels used for training were restricted to $2.5 \lesssim z_{\rm abs} \lesssim 2.7$, the UVES spectra were drawn from quasars with $2.5 \lesssim z_{\rm QSO} \lesssim 3.0$, and therefore contained Ly$\alpha$ absorption over a broader redshift range. We further examined the prediction accuracy as a function of absorber redshift and found that the MAE remains nearly constant across $2.0 \lesssim z_{\rm abs} \lesssim 3.0$, with mean values of $0.328 \pm 0.022$ for $\log N_{\rm HI}$ and $0.193 \pm 0.011$ for $\log b_{\rm HI}$ across the redshift bins. No significant trend is observed for $\log b_{\rm HI}$, while $\log N_{\rm HI}$ shows only a weak decrease in MAE toward higher redshift. 
These results indicate that the CNN can successfully reproduce the absorber parameters identified by the fitting procedure beyond the specific absorber redshift interval used during training.

Our findings are broadly consistent with previous machine-learning approaches developed for Ly$\alpha$ absorber identification and parameter estimation. \citet{Cheng2022MNRAS.517..755C} developed a CNN-based framework to identify low-column-density Ly$\alpha$ absorbers and predict their physical parameters, including $\log N_{\rm HI}$, absorber redshift, and Doppler width. Applied to simulated spectra, they found that more than $99\%$ of CNN detections corresponded to true systems, although the completeness was initially lower due to missed weak and blended absorption features. When applied to high-resolution HIRES quasar spectra, approximately $78\%$ of CNN-identified systems were recovered in manual Voigt-profile fitting catalogs. They further demonstrated reliable parameter recovery in the range $12.5\leq\log N_{\rm HI}<15.5$, obtaining $\mathrm{MAE}\simeq0.13$ in $\log N_{\rm HI}$ and $\Delta b_{\rm HI}\simeq4.1~{\rm km~s^{-1}}$. A key result of their study was that the CNN predictions remained stable over a broad range of S/N values, suggesting that such models can be applied efficiently to both current and future spectroscopic surveys. Importantly, however, they also reported a degradation in performance when moving from simulated to observed spectra, attributing it primarily to weak absorbers, line blending, and complexities present in real observations that are difficult to capture in simulations.
Similarly, \citet{Jalan2024A&A...688A.126J} trained on $\sim10^6$ simulated low-redshift Ly$\alpha$ profiles designed to mimic HST-COS observations and achieved high performance on simulated data, correctly identifying more than $98\%$ of single-component systems and over $90\%$ of double-component systems. For $90\%$ of single absorbers, the predicted Doppler widths were recovered within $\sim\pm8~{\rm km\,s^{-1}}$ and $\log N_{\rm HI}$ within $\sim\pm0.3$, while somewhat larger uncertainties were obtained for blended double-component systems. The authors also showed that the performance improves with increasing S/N and remains particularly robust for systems with S/N$>20$. However, when applied to real HST-COS observations, the component-classification accuracy decreased by approximately $10\%$. Despite this reduction, the predicted $b$ and $N_{\rm HI}$ distributions remained in reasonable agreement with traditional Voigt-profile fitting approaches such as VIPER and \cite{Danforth2016ApJ...817..111D}. Through additional mock-data tests, the authors concluded that the primary origin of the simulation-to-observation performance gap is the inability of simulated training samples to fully reproduce the complexities of real spectra, including realistic noise characteristics, instrumental effects, and subtle observational features.
A similar trend is observed in our analysis. While the CNN performs well on simulated spectra, its performance decreases for real UVES observations, with the F1 score declining from $\sim0.8$ for simulated data to $\sim0.5$ for real spectra. This behaviour suggests that the simulation-to-observation domain shift remains a common challenge for machine-learning-based analyses of Ly$\alpha$ absorption systems.
Nevertheless, as also emphasized by previous studies, the principal advantage of CNN-based methods lies in their computational efficiency. For example, \citet{Cheng2022MNRAS.517..755C} demonstrated that CNN-based approaches can provide rapid predictions for large Ly$\alpha$ forest spectra, substantially reducing the computational burden associated with traditional Voigt-profile analyses. Once trained, these models provide rapid predictions that can facilitate the analysis of large spectral datasets. In addition, our work extends previous approaches by incorporating a Monte Carlo dropout framework during inference to estimate predictive uncertainties. This allows the model not only to provide parameter predictions but also to assign confidence estimates, potentially leading to more reliable characterization of absorbers, particularly in blended or complex systems.
\section{Summary}
\label{sec: summary}
The main results of this work are summarized below:

\begin{itemize}

\item We developed a sliding-window convolutional neural network (CNN) to identify Ly$\alpha$ absorbers and predict their properties ($N_{\rm HI}$, $b_{\rm HI}$ and absorber location) directly from quasar spectra.

\item Monte Carlo dropout enables estimation of predictive uncertainties during inference, providing confidence measures alongside absorber parameter predictions. However, these uncertainties should be interpreted with caution, as they are not fully calibrated total uncertainties, particularly for $\log b_{\rm HI}$ in observational spectra.

\item The network was trained on synthetic Ly$\alpha$ forest spectra generated from the IllustrisTNG simulation, with absorber labels obtained from Voigt-profile fitting using \textsc{viper}.

\item For simulated spectra, the CNN achieves an F1 score of $\sim0.8$ when evaluated against the \textsc{viper} absorber catalogue, with $\mathrm{MAE}=0.18$ and $\mathrm{RMSE}=0.27$ for $\log N_{\rm HI}$, and $\mathrm{MAE}=0.10$ and $\mathrm{RMSE}=0.17$ for $\log b_{\rm HI}$.

\item Prediction accuracy improves with increasing SNR and reaches its best performance for intermediate-to-high quality spectra ($\mathrm{SNR}\sim30$--40).

\item The CNN accurately reproduces key Ly$\alpha$ forest statistics characterized by \textsc{viper} in simulations, recovering a CDDF slope of $\beta=1.68^{+0.06}_{-0.12}$ compared to $\beta=1.73^{+0.07}_{-0.05}$ from \textsc{viper}, and reproducing the lower envelope of the $b_{\rm HI}$--$N_{\rm HI}$ relation with an RMS difference of only $0.36~{\rm km~s^{-1}}$.

\item When applied to real UVES spectra, performance decreases, yielding an F1 score of $\sim0.5$, $\mathrm{MAE}=0.35$ and $\mathrm{RMSE}=0.50$ for $\\log N_{\rm HI}$, and $\mathrm{MAE}=0.22$ and $\mathrm{RMSE}=0.34$ for $\log b_{\rm HI}$, when evaluated against the \textsc{viper} absorber catalogue.

\item Analysis of the learned latent representations reveals a significant domain shift between the simulated training data and the observational spectra, indicating that this mismatch is a major contributor to the reduced performance on real data.

\item Despite this degradation, the CNN preserves the global statistical properties of the observed Ly$\alpha$ forest as characterized by \textsc{viper}, recovering a CDDF slope of $\beta=1.36^{+0.10}_{-0.10}$ compared to $\beta=1.40^{+0.06}_{-0.04}$ from \textsc{viper}, while reproducing the lower-envelope relation with an RMS difference of $3.20\,{\rm km\,s^{-1}}$.

\item The model generalizes successfully beyond the absorber redshift range used for training, indicating that its predictions are not restricted to the specific redshift interval represented in the training sample.

\item Once trained, the CNN processes a UVES spectrum approximately 4.3 times faster than \textsc{viper} when using 50 Monte Carlo dropout realizations on an NVIDIA T4 GPU, demonstrating its computational efficiency for high-throughput spectroscopic analysis.

\end{itemize}

In summary, we demonstrate that a simulation-trained sliding-window CNN can rapidly identify and characterize Ly$\alpha$ absorbers while reproducing the key statistical properties of the Ly$\alpha$ forest as characterized by traditional Voigt-profile fitting. Although performance degrades when applied to real spectra, the model remains capable of reproducing absorber populations and distributions consistent with those obtained from traditional Voigt-profile fitting. Future work should focus on reducing the domain gap between simulations and observations through more realistic mock spectra that incorporate instrumental and observational effects, together with domain-adaptation techniques, to further improve performance on real data. These results highlight the potential of deep-learning approaches as fast and scalable tools for future large spectroscopic surveys.

\section*{Acknowledgements}
The research of P.S. is supported by the University Grants Commission (UGC), Government of India, under the UGC-JRF scheme  (Ref. No.: 221610014755). H.C. and P.S. express their gratitude to the Inter-University Centre for Astronomy and Astrophysics (IUCAA) for their hospitality and the provision of High-Performance Computing (HPC) facilities under the IUCAA Associate Programme. The authors would like to thank Dr. Sukanya Mallik for valuable discussions on the generation of synthetic spectra from cosmological simulations, and Dr. Ninan Sajeeth Philip for insightful discussions on convolutional neural network architectures and their implementation.
Additionally, we acknowledge the assistance of AI tools, specifically OpenAI's ChatGPT, for aiding in writing and code development, and Grammarly for enhancing the text's clarity and correctness.
\section*{Data Availability}
The data used in this study are publicly available through the IllustrisTNG data release and the UVES spectral data release. The machine-learning code developed for this work is not publicly available but may be obtained from the corresponding author upon reasonable request.


\bibliographystyle{mnras}
\bibliography{biblography} 

@ARTICLE{Nelson2019ComAC...6....2N,
       author = {{Nelson}, Dylan and {Springel}, Volker and {Pillepich}, Annalisa and {Rodriguez-Gomez}, Vicente and {Torrey}, Paul and {Genel}, Shy and {Vogelsberger}, Mark and {Pakmor}, Ruediger and {Marinacci}, Federico and {Weinberger}, Rainer and {Kelley}, Luke and {Lovell}, Mark and {Diemer}, Benedikt and {Hernquist}, Lars},
        title = "{The IllustrisTNG simulations: public data release}",
      journal = {Computational Astrophysics and Cosmology},
         year = 2019,
        month = may,
       volume = {6},
       number = {1},
          eid = {2},
        pages = {2},
          doi = {10.1186/s40668-019-0028-x},
archivePrefix = {arXiv},
       eprint = {1812.05609},
 primaryClass = {astro-ph.GA},
       adsurl = {https://ui.adsabs.harvard.edu/abs/2019ComAC...6....2N}
}

@ARTICLE{Hummels2017ApJ...847...59H,
       author = {{Hummels}, Cameron B. and {Smith}, Britton D. and {Silvia}, Devin W.},
        title = "{Trident: A Universal Tool for Generating Synthetic Absorption Spectra from Astrophysical Simulations}",
      journal = {\apj},
         year = 2017,
        month = sep,
       volume = {847},
       number = {1},
          eid = {59},
        pages = {59},
          doi = {10.3847/1538-4357/aa7e2d},
archivePrefix = {arXiv},
       eprint = {1612.03935},
 primaryClass = {astro-ph.IM},
       adsurl = {https://ui.adsabs.harvard.edu/abs/2017ApJ...847...59H}
}

@ARTICLE{Gaikwad2017MNRAS.467.3172G,
       author = {{Gaikwad}, Prakash and {Srianand}, Raghunathan and {Choudhury}, Tirthankar Roy and {Khaire}, Vikram},
        title = "{VoIgt profile Parameter Estimation Routine (viper): H I photoionization rate at z < 0.5}",
      journal = {\mnras},
         year = 2017,
        month = may,
       volume = {467},
       number = {3},
        pages = {3172-3187},
          doi = {10.1093/mnras/stx248},
archivePrefix = {arXiv},
       eprint = {1610.06572},
 primaryClass = {astro-ph.CO},
       adsurl = {https://ui.adsabs.harvard.edu/abs/2017MNRAS.467.3172G}
}

@ARTICLE{Cheng2022MNRAS.517..755C,
       author = {{Cheng}, Ting-Yun and {Cooke}, Ryan J. and {Rudie}, Gwen},
        title = "{Harvesting the Ly {\ensuremath{\alpha}} forest with convolutional neural networks}",
      journal = {\mnras},
         year = 2022,
        month = nov,
       volume = {517},
       number = {1},
        pages = {755-775},
          doi = {10.1093/mnras/stac2631},
archivePrefix = {arXiv},
       eprint = {2209.02142},
 primaryClass = {astro-ph.GA},
       adsurl = {https://ui.adsabs.harvard.edu/abs/2022MNRAS.517..755C}
}

@ARTICLE{Bainbridge2017MNRAS.468.1639B,
       author = {{Bainbridge}, Matthew B. and {Webb}, John K.},
        title = "{Artificial intelligence applied to the automatic analysis of absorption spectra. Objective measurement of the fine structure constant}",
      journal = {\mnras},
         year = 2017,
        month = jun,
       volume = {468},
       number = {2},
        pages = {1639-1670},
          doi = {10.1093/mnras/stx179},
archivePrefix = {arXiv},
       eprint = {1606.07393},
 primaryClass = {astro-ph.IM},
       adsurl = {https://ui.adsabs.harvard.edu/abs/2017MNRAS.468.1639B}
}

@ARTICLE{Meiksin2009RvMP...81.1405M,
       author = {{Meiksin}, Avery A.},
        title = "{The physics of the intergalactic medium}",
      journal = {Reviews of Modern Physics},
         year = 2009,
        month = oct,
       volume = {81},
       number = {4},
        pages = {1405-1469},
          doi = {10.1103/RevModPhys.81.1405},
archivePrefix = {arXiv},
       eprint = {0711.3358},
 primaryClass = {astro-ph},
       adsurl = {https://ui.adsabs.harvard.edu/abs/2009RvMP...81.1405M}
}

@ARTICLE{Kingma2014arXiv1412.6980K,
       author = {{Kingma}, Diederik P. and {Ba}, Jimmy},
        title = "{Adam: A Method for Stochastic Optimization}",
      journal = {arXiv e-prints},
         year = 2014,
        month = dec,
          eid = {arXiv:1412.6980},
        pages = {arXiv:1412.6980},
          doi = {10.48550/arXiv.1412.6980},
archivePrefix = {arXiv},
       eprint = {1412.6980},
 primaryClass = {cs.LG},
       adsurl = {https://ui.adsabs.harvard.edu/abs/2014arXiv1412.6980K}
}

@ARTICLE{Rauch1998ARA&A..36..267R,
       author = {{Rauch}, Michael},
        title = "{The Lyman Alpha Forest in the Spectra of QSOs}",
      journal = {\araa},
         year = 1998,
        month = jan,
       volume = {36},
        pages = {267-316},
          doi = {10.1146/annurev.astro.36.1.267},
archivePrefix = {arXiv},
       eprint = {astro-ph/9806286},
 primaryClass = {astro-ph},
       adsurl = {https://ui.adsabs.harvard.edu/abs/1998ARA&A..36..267R}
}

@ARTICLE{Lynds1971ApJ...164L..73L,
       author = {{Lynds}, Roger},
        title = "{The Absorption-Line Spectrum of 4c 05.34}",
      journal = {\apjl},
         year = 1971,
        month = mar,
       volume = {164},
        pages = {L73},
          doi = {10.1086/180695},
       adsurl = {https://ui.adsabs.harvard.edu/abs/1971ApJ...164L..73L}
}

@ARTICLE{Schaye2001ApJ...559..507S,
       author = {{Schaye}, Joop},
        title = "{Model-independent Insights into the Nature of the Ly{\ensuremath{\alpha}} Forest and the Distribution of Matter in the Universe}",
      journal = {\apj},
         year = 2001,
        month = oct,
       volume = {559},
       number = {2},
        pages = {507-515},
          doi = {10.1086/322421},
archivePrefix = {arXiv},
       eprint = {astro-ph/0104272},
 primaryClass = {astro-ph},
       adsurl = {https://ui.adsabs.harvard.edu/abs/2001ApJ...559..507S}
}

@ARTICLE{Becker2013MNRAS.430.2067B,
       author = {{Becker}, George D. and {Hewett}, Paul C. and {Worseck}, G{\'a}bor and {Prochaska}, J. Xavier},
        title = "{A refined measurement of the mean transmitted flux in the Ly{\ensuremath{\alpha}} forest over 2 < z < 5 using composite quasar spectra}",
      journal = {\mnras},
         year = 2013,
        month = apr,
       volume = {430},
       number = {3},
        pages = {2067-2081},
          doi = {10.1093/mnras/stt031},
archivePrefix = {arXiv},
       eprint = {1208.2584},
 primaryClass = {astro-ph.CO},
       adsurl = {https://ui.adsabs.harvard.edu/abs/2013MNRAS.430.2067B}
}

@ARTICLE{Viel2013MNRAS.429.1734V,
       author = {{Viel}, Matteo and {Schaye}, Joop and {Booth}, C.~M.},
        title = "{The impact of feedback from galaxy formation on the Lyman {\ensuremath{\alpha}} transmitted flux}",
      journal = {\mnras},
         year = 2013,
        month = feb,
       volume = {429},
       number = {2},
        pages = {1734-1746},
          doi = {10.1093/mnras/sts465},
archivePrefix = {arXiv},
       eprint = {1207.6567},
 primaryClass = {astro-ph.CO},
       adsurl = {https://ui.adsabs.harvard.edu/abs/2013MNRAS.429.1734V}
}

@ARTICLE{Irsic2017PhRvD..96b3522I,
       author = {{Ir{\v{s}}i{\v{c}}}, Vid and {Viel}, Matteo and {Haehnelt}, Martin G. and {Bolton}, James S. and {Cristiani}, Stefano and {Becker}, George D. and {D'Odorico}, Valentina and {Cupani}, Guido and {Kim}, Tae-Sun and {Berg}, Trystyn A.~M. and {L{\'o}pez}, Sebastian and {Ellison}, Sara and {Christensen}, Lise and {Denney}, Kelly D. and {Worseck}, G{\'a}bor},
        title = "{New constraints on the free-streaming of warm dark matter from intermediate and small scale Lyman-{\ensuremath{\alpha}} forest data}",
      journal = {\prd},
         year = 2017,
        month = jul,
       volume = {96},
       number = {2},
          eid = {023522},
        pages = {023522},
          doi = {10.1103/PhysRevD.96.023522},
archivePrefix = {arXiv},
       eprint = {1702.01764},
 primaryClass = {astro-ph.CO},
       adsurl = {https://ui.adsabs.harvard.edu/abs/2017PhRvD..96b3522I}
}

@ARTICLE{Rogers2021PhRvL.126g1302R,
       author = {{Rogers}, Keir K. and {Peiris}, Hiranya V.},
        title = "{Strong Bound on Canonical Ultralight Axion Dark Matter from the Lyman-Alpha Forest}",
      journal = {\prl},
         year = 2021,
        month = feb,
       volume = {126},
       number = {7},
          eid = {071302},
        pages = {071302},
          doi = {10.1103/PhysRevLett.126.071302},
archivePrefix = {arXiv},
       eprint = {2007.12705},
 primaryClass = {astro-ph.CO},
       adsurl = {https://ui.adsabs.harvard.edu/abs/2021PhRvL.126g1302R}
}

@ARTICLE{Cristiani1995MNRAS.273.1016C,
       author = {{Cristiani}, Stefano and {D'Odorico}, Sandro and {Fontana}, Adriano and {Giallongo}, Emanuele and {Savaglio}, Sandra},
        title = "{The space distribution of the Lyman alpha clouds in the line of sight to the z=3.66 QSO 0055-269}",
      journal = {\mnras},
         year = 1995,
        month = apr,
       volume = {273},
       number = {4},
        pages = {1016-1032},
          doi = {10.1093/mnras/273.4.1016},
archivePrefix = {arXiv},
       eprint = {astro-ph/9411075},
 primaryClass = {astro-ph},
       adsurl = {https://ui.adsabs.harvard.edu/abs/1995MNRAS.273.1016C}
}

@ARTICLE{Fang1996ApJ...462...77F,
       author = {{Fang}, Yihu and {Duncan}, Robert C. and {Crotts}, Arlin P.~S. and {Bechtold}, Jill},
        title = "{The Size and Nature of Lyman- alpha Forest Clouds Probed by QSO Pairs and Groups}",
      journal = {\apj},
         year = 1996,
        month = may,
       volume = {462},
        pages = {77},
          doi = {10.1086/177129},
archivePrefix = {arXiv},
       eprint = {astro-ph/9510112},
 primaryClass = {astro-ph},
       adsurl = {https://ui.adsabs.harvard.edu/abs/1996ApJ...462...77F}
}

@ARTICLE{Pedersen2023ApJ...944..223P,
       author = {{Pedersen}, Christian and {Font-Ribera}, Andreu and {Gnedin}, Nickolay Y.},
        title = "{Compressing the Cosmological Information in One-dimensional Correlations of the Lyman-{\ensuremath{\alpha}} Forest}",
      journal = {\apj},
         year = 2023,
        month = feb,
       volume = {944},
       number = {2},
          eid = {223},
        pages = {223},
          doi = {10.3847/1538-4357/acb433},
archivePrefix = {arXiv},
       eprint = {2209.09895},
 primaryClass = {astro-ph.CO},
       adsurl = {https://ui.adsabs.harvard.edu/abs/2023ApJ...944..223P}
}

@software{Carswell2014ascl.soft08015C,
       author = {{Carswell}, R.~F. and {Webb}, J.~K.},
        title = "{VPFIT: Voigt profile fitting program}",
 howpublished = {Astrophysics Source Code Library, record ascl:1408.015},
         year = 2014,
        month = aug,
          eid = {ascl:1408.015},
archivePrefix = {ascl},
       eprint = {1408.015},
       adsurl = {https://ui.adsabs.harvard.edu/abs/2014ascl.soft08015C}
}

@ARTICLE{Liang2017arXiv171009852L,
       author = {{Liang}, Cameron and {Kravtsov}, Andrey},
        title = "{BayesVP: a Bayesian Voigt profile fitting package}",
      journal = {arXiv e-prints},
         year = 2017,
        month = oct,
          eid = {arXiv:1710.09852},
        pages = {arXiv:1710.09852},
          doi = {10.48550/arXiv.1710.09852},
archivePrefix = {arXiv},
       eprint = {1710.09852},
 primaryClass = {astro-ph.GA},
       adsurl = {https://ui.adsabs.harvard.edu/abs/2017arXiv171009852L}
}

@ARTICLE{Krogager2018arXiv180301187K,
       author = {{Krogager}, Jens-Kristian},
        title = "{VoigtFit: A Python package for Voigt profile fitting}",
      journal = {arXiv e-prints},
         year = 2018,
        month = mar,
          eid = {arXiv:1803.01187},
        pages = {arXiv:1803.01187},
          doi = {10.48550/arXiv.1803.01187},
archivePrefix = {arXiv},
       eprint = {1803.01187},
 primaryClass = {astro-ph.IM},
       adsurl = {https://ui.adsabs.harvard.edu/abs/2018arXiv180301187K}
}

@ARTICLE{DESI2025arXiv250314745D,
       author = {{DESI Collaboration} and {Abdul-Karim}, M. and {Adame}, A.~G. and {Aguado}, D. and {Aguilar}, J. and {Ahlen}, S. and {Alam}, S. and {Aldering}, G. and {Alexander}, D.~M. and {Alfarsy}, R. and {Allen}, L. and {Allende Prieto}, C. and {Alves}, O. and {Anand}, A. and {Andrade}, U. and {Armengaud}, E. and {Avila}, S. and {Aviles}, A. and {Awan}, H. and {Bailey}, S. and {Baleato Lizancos}, A. and {Ballester}, O. and {Bault}, A. and {Bautista}, J. and {BenZvi}, S. and {Beraldo e Silva}, L. and {Bermejo-Climent}, J.~R. and {Beutler}, F. and {Bianchi}, D. and {Blake}, C. and {Blum}, R. and {Bolton}, A.~S. and {Bonici}, M. and {Brieden}, S. and {Brodzeller}, A. and {Brooks}, D. and {Buckley-Geer}, E. and {Burtin}, E. and {Canning}, R. and {Carnero Rosell}, A. and {Carr}, A. and {Carrilho}, P. and {Casas}, L. and {Castander}, F.~J. and {Cereskaite}, R. and {Cervantes-Cota}, J.~L. and {Chaussidon}, E. and {Chaves-Montero}, J. and {Chen}, S. and {Chen}, X. and {Claybaugh}, T. and {Cole}, S. and {Cooper}, A.~P. and {Cousinou}, M.-C. and {Cuceu}, A. and {Davis}, T.~M. and {Dawson}, K.~S. and {de Belsunce}, R. and {de la Cruz}, R. and {de la Macorra}, A. and {de Mattia}, A. and {Deiosso}, N. and {Della Costa}, J. and {Demina}, R. and {Demirbozan}, U. and {DeRose}, J. and {Dey}, A. and {Dey}, B. and {Ding}, J. and {Ding}, Z. and {Doel}, P. and {Douglass}, K. and {Dowicz}, M. and {Ebina}, H. and {Edelstein}, J. and {Eisenstein}, D.~J. and {Elbers}, W. and {Emas}, N. and {Escoffier}, S. and {Fagrelius}, P. and {Fan}, X. and {Fanning}, K. and {Fawcett}, V.~A. and {Fern\textbackslash'andez-Garc\textbackslash'ia}, E. and {Ferraro}, S. and {Findlay}, N. and {Font-Ribera}, A. and {Forero-Romero}, J.~E. and {Forero-S\textbackslash'anchez}, D. and {Frenk}, C.~S. and {G\textbackslash''ansicke}, B.~T. and {Galbany}, L. and {Garc\textbackslash'ia-Bellido}, J. and {Garcia-Quintero}, C. and {Garrison}, L.~H. and {Gazta\textbackslash\raisebox{-0.5ex}\textasciitildenaga}, E. and {Gil-Mar\textbackslash'in}, H. and {Gnedin}, O.~Y. and {Gontcho}, S. Gontcho A and {Gonzalez-Morales}, A.~X. and {Gonzalez-Perez}, V. and {Gordon}, C. and {Graur}, O. and {Green}, D. and {Gruen}, D. and {Gsponer}, R. and {Guandalin}, C. and {Gutierrez}, G. and {Guy}, J. and {Hahn}, C. and {Han}, J.~J. and {Han}, J. and {He}, S. and {Herrera-Alcantar}, H.~K. and {Honscheid}, K. and {Hou}, J. and {Howlett}, C. and {Huterer}, D. and {Ir\textbackslashv\{s\}i\textbackslashv\{c\}}, V. and {Ishak}, M. and {Jacques}, A. and {Jimenez}, J. and {Jing}, Y.~P. and {Joachimi}, B. and {Joudaki}, S. and {Joyce}, R. and {Jullo}, E. and {Juneau}, S. and {Kara\textbackslashc\{c\}ayl\{\textbackslashi\}}, N.~G. and {Karim}, T. and {Kehoe}, R. and {Kent}, S. and {Khederlarian}, A. and {Kirkby}, D. and {Kisner}, T. and {Kitaura}, F.-S. and {Kizhuprakkat}, N. and {Kong}, H. and {Koposov}, S.~E. and {Kremin}, A. and {Krolewski}, A. and {Lahav}, O. and {Lai}, Y. and {Lamman}, C. and {Lan}, T.-W. and {Landriau}, M. and {Lang}, D. and {Lange}, J.~U. and {Lasker}, J. and {Le Goff}, J.~M. and {Le Guillou}, L. and {Leauthaud}, A. and {Levi}, M.~E. and {Li}, S. and {Li}, T.~S. and {Lodha}, K. and {Lokken}, M. and {Luo}, Y. and {Magneville}, C. and {Manera}, M. and {Manser}, C.~J. and {Margala}, D. and {Martini}, P. and {Maus}, M. and {McCullough}, J. and {McDonald}, P. and {Medina}, G.~E. and {Medina-Varela}, L. and {Meisner}, A. and {Mena-Fern\textbackslash'andez}, J. and {Menegas}, A. and {Mezcua}, M. and {Miquel}, R. and {Montero-Camacho}, P. and {Moon}, J. and {Moustakas}, J. and {Mu\textbackslash\raisebox{-0.5ex}\textasciitildenoz-Guti\textbackslash'errez}, A. and {Mu\textbackslash\raisebox{-0.5ex}\textasciitildenoz-Santos}, D. and {Myers}, A.~D. and {Myles}, J. and {Nadathur}, S. and {Najita}, J. and {Napolitano}, L. and {Newman}, J.~A. and {Nikakhtar}, F. and {Nikutta}, R. and {Niz}, G. and {Noriega}, H.~E. and {Padmanabhan}, N. and {Paillas}, E. and {Palanque-Delabrouille}, N. and {Palmese}, A. and {Pan}, J. and {Pan}, Z. and {Parkinson}, D. and {Peacock}, J. and {Percival}, W.~J. and {P\textbackslash'erez-Fern\textbackslash'andez}, A. and {P\textbackslash'erez-R\textbackslash`afols}, I. and {Peterson}, P.},
        title = "{Data Release 1 of the Dark Energy Spectroscopic Instrument}",
      journal = {arXiv e-prints},
         year = 2025,
        month = mar,
          eid = {arXiv:2503.14745},
        pages = {arXiv:2503.14745},
          doi = {10.48550/arXiv.2503.14745},
archivePrefix = {arXiv},
       eprint = {2503.14745},
 primaryClass = {astro-ph.CO},
       adsurl = {https://ui.adsabs.harvard.edu/abs/2025arXiv250314745D}
}

@ARTICLE{Cheng2020MNRAS.493.4209C,
       author = {{Cheng}, Ting-Yun and {Conselice}, Christopher J. and {Arag{\'o}n-Salamanca}, Alfonso and {Li}, Nan and {Bluck}, Asa F.~L. and {Hartley}, Will G. and {Annis}, James and {Brooks}, David and {Doel}, Peter and {Garc{\'\i}a-Bellido}, Juan and {James}, David J. and {Kuehn}, Kyler and {Kuropatkin}, Nikolay and {Smith}, Mathew and {Sobreira}, Flavia and {Tarle}, Gregory},
        title = "{Optimizing automatic morphological classification of galaxies with machine learning and deep learning using Dark Energy Survey imaging}",
      journal = {\mnras},
         year = 2020,
        month = apr,
       volume = {493},
       number = {3},
        pages = {4209-4228},
          doi = {10.1093/mnras/staa501},
archivePrefix = {arXiv},
       eprint = {1908.03610},
 primaryClass = {astro-ph.GA},
       adsurl = {https://ui.adsabs.harvard.edu/abs/2020MNRAS.493.4209C}
}

@ARTICLE{Cheng2021MNRAS.507.4425C,
       author = {{Cheng}, Ting-Yun and {Conselice}, Christopher J. and {Arag{\'o}n-Salamanca}, Alfonso and {Aguena}, M. and {Allam}, S. and {Andrade-Oliveira}, F. and {Annis}, J. and {Bluck}, A.~F.~L. and {Brooks}, D. and {Burke}, D.~L. and {Carrasco Kind}, M. and {Carretero}, J. and {Choi}, A. and {Costanzi}, M. and {da Costa}, L.~N. and {Pereira}, M.~E.~S. and {De Vicente}, J. and {Diehl}, H.~T. and {Drlica-Wagner}, A. and {Eckert}, K. and {Everett}, S. and {Evrard}, A.~E. and {Ferrero}, I. and {Fosalba}, P. and {Frieman}, J. and {Garc{\'\i}a-Bellido}, J. and {Gerdes}, D.~W. and {Giannantonio}, T. and {Gruen}, D. and {Gruendl}, R.~A. and {Gschwend}, J. and {Gutierrez}, G. and {Hinton}, S.~R. and {Hollowood}, D.~L. and {Honscheid}, K. and {James}, D.~J. and {Krause}, E. and {Kuehn}, K. and {Kuropatkin}, N. and {Lahav}, O. and {Maia}, M.~A.~G. and {March}, M. and {Menanteau}, F. and {Miquel}, R. and {Morgan}, R. and {Paz-Chinch{\'o}n}, F. and {Pieres}, A. and {Plazas Malag{\'o}n}, A.~A. and {Roodman}, A. and {Sanchez}, E. and {Scarpine}, V. and {Serrano}, S. and {Sevilla-Noarbe}, I. and {Smith}, M. and {Soares-Santos}, M. and {Suchyta}, E. and {Swanson}, M.~E.~C. and {Tarle}, G. and {Thomas}, D. and {To}, C.},
        title = "{Galaxy morphological classification catalogue of the Dark Energy Survey Year 3 data with convolutional neural networks}",
      journal = {\mnras},
         year = 2021,
        month = nov,
       volume = {507},
       number = {3},
        pages = {4425-4444},
          doi = {10.1093/mnras/stab2142},
archivePrefix = {arXiv},
       eprint = {2107.10210},
 primaryClass = {astro-ph.GA},
       adsurl = {https://ui.adsabs.harvard.edu/abs/2021MNRAS.507.4425C}
}

@ARTICLE{Walmsley2022MNRAS.509.3966W,
       author = {{Walmsley}, Mike and {Lintott}, Chris and {G{\'e}ron}, Tobias and {Kruk}, Sandor and {Krawczyk}, Coleman and {Willett}, Kyle W. and {Bamford}, Steven and {Kelvin}, Lee S. and {Fortson}, Lucy and {Gal}, Yarin and {Keel}, William and {Masters}, Karen L. and {Mehta}, Vihang and {Simmons}, Brooke D. and {Smethurst}, Rebecca and {Smith}, Lewis and {Baeten}, Elisabeth M. and {Macmillan}, Christine},
        title = "{Galaxy Zoo DECaLS: Detailed visual morphology measurements from volunteers and deep learning for 314 000 galaxies}",
      journal = {\mnras},
         year = 2022,
        month = jan,
       volume = {509},
       number = {3},
        pages = {3966-3988},
          doi = {10.1093/mnras/stab2093},
archivePrefix = {arXiv},
       eprint = {2102.08414},
 primaryClass = {astro-ph.GA},
       adsurl = {https://ui.adsabs.harvard.edu/abs/2022MNRAS.509.3966W}
}

@ARTICLE{Metcalf2019A&A...625A.119M,
       author = {{Metcalf}, R.~B. and {Meneghetti}, M. and {Avestruz}, C. and {Bellagamba}, F. and {Bom}, C.~R. and {Bertin}, E. and {Cabanac}, R. and {Courbin}, F. and {Davies}, A. and {Decenci{\`e}re}, E. and {Flamary}, R. and {Gavazzi}, R. and {Geiger}, M. and {Hartley}, P. and {Huertas-Company}, M. and {Jackson}, N. and {Jacobs}, C. and {Jullo}, E. and {Kneib}, J.-P. and {Koopmans}, L.~V.~E. and {Lanusse}, F. and {Li}, C.-L. and {Ma}, Q. and {Makler}, M. and {Li}, N. and {Lightman}, M. and {Petrillo}, C.~E. and {Serjeant}, S. and {Sch{\"a}fer}, C. and {Sonnenfeld}, A. and {Tagore}, A. and {Tortora}, C. and {Tuccillo}, D. and {Valent{\'\i}n}, M.~B. and {Velasco-Forero}, S. and {Verdoes Kleijn}, G.~A. and {Vernardos}, G.},
        title = "{The strong gravitational lens finding challenge}",
      journal = {\aap},
         year = 2019,
        month = may,
       volume = {625},
          eid = {A119},
        pages = {A119},
          doi = {10.1051/0004-6361/201832797},
archivePrefix = {arXiv},
       eprint = {1802.03609},
 primaryClass = {astro-ph.GA},
       adsurl = {https://ui.adsabs.harvard.edu/abs/2019A&A...625A.119M}
}

@ARTICLE{Muthukrishna2019PASP..131k8002M,
       author = {{Muthukrishna}, Daniel and {Narayan}, Gautham and {Mandel}, Kaisey S. and {Biswas}, Rahul and {Hlo{\v{z}}ek}, Ren{\'e}e},
        title = "{RAPID: Early Classification of Explosive Transients Using Deep Learning}",
      journal = {\pasp},
         year = 2019,
        month = nov,
       volume = {131},
       number = {1005},
        pages = {118002},
          doi = {10.1088/1538-3873/ab1609},
archivePrefix = {arXiv},
       eprint = {1904.00014},
 primaryClass = {astro-ph.IM},
       adsurl = {https://ui.adsabs.harvard.edu/abs/2019PASP..131k8002M}
}

@ARTICLE{Garnett2017MNRAS.472.1850G,
       author = {{Garnett}, Roman and {Ho}, Shirley and {Bird}, Simeon and {Schneider}, Jeff},
        title = "{Detecting damped Ly {\ensuremath{\alpha}} absorbers with Gaussian processes}",
      journal = {\mnras},
         year = 2017,
        month = dec,
       volume = {472},
       number = {2},
        pages = {1850-1865},
          doi = {10.1093/mnras/stx1958},
archivePrefix = {arXiv},
       eprint = {1605.04460},
 primaryClass = {astro-ph.CO},
       adsurl = {https://ui.adsabs.harvard.edu/abs/2017MNRAS.472.1850G}
}

@ARTICLE{Parks2018MNRAS.476.1151P,
       author = {{Parks}, David and {Prochaska}, J. Xavier and {Dong}, Shawfeng and {Cai}, Zheng},
        title = "{Deep learning of quasar spectra to discover and characterize damped Ly{\ensuremath{\alpha}} systems}",
      journal = {\mnras},
         year = 2018,
        month = may,
       volume = {476},
       number = {1},
        pages = {1151-1168},
          doi = {10.1093/mnras/sty196},
archivePrefix = {arXiv},
       eprint = {1709.04962},
 primaryClass = {astro-ph.GA},
       adsurl = {https://ui.adsabs.harvard.edu/abs/2018MNRAS.476.1151P}
}

@ARTICLE{Huang2021MNRAS.506.5212H,
       author = {{Huang}, Lawrence and {Croft}, Rupert A.~C. and {Arora}, Hitesh},
        title = "{Deep forest: Neural network reconstruction of the Lyman-{\ensuremath{\alpha}} forest}",
      journal = {\mnras},
         year = 2021,
        month = oct,
       volume = {506},
       number = {4},
        pages = {5212-5222},
          doi = {10.1093/mnras/stab2041},
archivePrefix = {arXiv},
       eprint = {2009.10673},
 primaryClass = {astro-ph.CO},
       adsurl = {https://ui.adsabs.harvard.edu/abs/2021MNRAS.506.5212H}
}

@ARTICLE{Veiga2021arXiv210709082V,
       author = {{Veiga}, Maria Han and {Meng}, Xi and {Gnedin}, Oleg Y. and {Gnedin}, Nickolay Y. and {Huan}, Xun},
        title = "{Reconstruction of the Density Power Spectrum from Quasar Spectra using Machine Learning}",
      journal = {arXiv e-prints},
         year = 2021,
        month = jul,
          eid = {arXiv:2107.09082},
        pages = {arXiv:2107.09082},
          doi = {10.48550/arXiv.2107.09082},
archivePrefix = {arXiv},
       eprint = {2107.09082},
 primaryClass = {astro-ph.CO},
       adsurl = {https://ui.adsabs.harvard.edu/abs/2021arXiv210709082V}
}

@ARTICLE{Stemock2024AJ....167..287S,
       author = {{Stemock}, Bryson and {Churchill}, Christopher W. and {Lee}, Avery and {Hassan}, Sultan and {Doughty}, Caitlin and {Ochoa}, Rogelio},
        title = "{Deep Learning Voigt Profiles. I. Single-Cloud Doublets}",
      journal = {\aj},
         year = 2024,
        month = jun,
       volume = {167},
       number = {6},
          eid = {287},
        pages = {287},
          doi = {10.3847/1538-3881/ad402b},
archivePrefix = {arXiv},
       eprint = {2311.16029},
 primaryClass = {astro-ph.GA},
       adsurl = {https://ui.adsabs.harvard.edu/abs/2024AJ....167..287S}
}

@ARTICLE{Pistis2025A&A...698A.292P,
       author = {{Pistis}, Francesco and {Fumagalli}, Michele and {Fossati}, Matteo and {Berg}, Trystyn and {Mangola}, Elena S. and {Dutta}, Rajeshwari and {Grespan}, Margherita and {Iovino}, Angela and {Ma{\l}ek}, Katarzyna and {Morrison}, Sean and {Murphy}, David N.~A. and {Pearson}, William J. and {P{\'e}rez-R{\'a}fols}, Ignasi and {Pieri}, Matthew M. and {Pollo}, Agnieszka and {Vergani}, Daniela},
        title = "{Automated quasar continuum estimation using neural networks: A comparative study of deep-learning architectures}",
      journal = {\aap},
         year = 2025,
        month = jun,
       volume = {698},
          eid = {A292},
        pages = {A292},
          doi = {10.1051/0004-6361/202453377},
archivePrefix = {arXiv},
       eprint = {2505.10976},
 primaryClass = {astro-ph.GA},
       adsurl = {https://ui.adsabs.harvard.edu/abs/2025A&A...698A.292P}
}

@ARTICLE{Sharma2020MNRAS.491.2280S,
       author = {{Sharma}, Kaushal and {Kembhavi}, Ajit and {Kembhavi}, Aniruddha and {Sivarani}, T. and {Abraham}, Sheelu and {Vaghmare}, Kaustubh},
        title = "{Application of convolutional neural networks for stellar spectral classification}",
      journal = {\mnras},
         year = 2020,
        month = jan,
       volume = {491},
       number = {2},
        pages = {2280-2300},
          doi = {10.1093/mnras/stz3100},
archivePrefix = {arXiv},
       eprint = {1909.05459},
 primaryClass = {astro-ph.SR},
       adsurl = {https://ui.adsabs.harvard.edu/abs/2020MNRAS.491.2280S}
}

@ARTICLE{Cabayol-Garcia2023MNRAS.525.3499C,
       author = {{Cabayol-Garcia}, L. and {Chaves-Montero}, J. and {Font-Ribera}, A. and {Pedersen}, C.},
        title = "{A neural network emulator for the Lyman-{\ensuremath{\alpha}} forest 1D flux power spectrum}",
      journal = {\mnras},
         year = 2023,
        month = nov,
       volume = {525},
       number = {3},
        pages = {3499-3515},
          doi = {10.1093/mnras/stad2512},
archivePrefix = {arXiv},
       eprint = {2305.19064},
 primaryClass = {astro-ph.CO},
       adsurl = {https://ui.adsabs.harvard.edu/abs/2023MNRAS.525.3499C}
}

@ARTICLE{Prochaska2014MNRAS.438..476P,
       author = {{Prochaska}, J. Xavier and {Madau}, Piero and {O'Meara}, John M. and {Fumagalli}, Michele},
        title = "{Towards a unified description of the intergalactic medium at redshift z {\ensuremath{\approx}} 2.5}",
      journal = {\mnras},
         year = 2014,
        month = feb,
       volume = {438},
       number = {1},
        pages = {476-486},
          doi = {10.1093/mnras/stt2218},
archivePrefix = {arXiv},
       eprint = {1310.0052},
 primaryClass = {astro-ph.CO},
       adsurl = {https://ui.adsabs.harvard.edu/abs/2014MNRAS.438..476P}
}

@ARTICLE{Hui1999ApJ...517..541H,
       author = {{Hui}, Lam and {Rutledge}, Robert E.},
        title = "{The b Distribution and the Velocity Structure of Absorption Peaks in the Ly{\ensuremath{\alpha}} Forest}",
      journal = {\apj},
         year = 1999,
        month = jun,
       volume = {517},
       number = {2},
        pages = {541-548},
          doi = {10.1086/307202},
archivePrefix = {arXiv},
       eprint = {astro-ph/9709100},
 primaryClass = {astro-ph},
       adsurl = {https://ui.adsabs.harvard.edu/abs/1999ApJ...517..541H}
}

@ARTICLE{Jalan2024A&A...688A.126J,
       author = {{Jalan}, P. and {Khaire}, V. and {Vivek}, M. and {Gaikwad}, P.},
        title = "{FLAME: Fitting Ly{\ensuremath{\alpha}} absorption lines using machine learning}",
      journal = {\aap},
         year = 2024,
        month = aug,
       volume = {688},
          eid = {A126},
        pages = {A126},
          doi = {10.1051/0004-6361/202449756},
archivePrefix = {arXiv},
       eprint = {2403.07498},
 primaryClass = {astro-ph.CO},
       adsurl = {https://ui.adsabs.harvard.edu/abs/2024A&A...688A.126J}
}

@ARTICLE{Danforth2016ApJ...817..111D,
       author = {{Danforth}, Charles W. and {Keeney}, Brian A. and {Tilton}, Evan M. and {Shull}, J. Michael and {Stocke}, John T. and {Stevans}, Matthew and {Pieri}, Matthew M. and {Savage}, Blair D. and {France}, Kevin and {Syphers}, David and {Smith}, Britton D. and {Green}, James C. and {Froning}, Cynthia and {Penton}, Steven V. and {Osterman}, Steven N.},
        title = "{An HST/COS Survey of the Low-redshift Intergalactic Medium. I. Survey, Methodology, and Overall Results}",
      journal = {\apj},
         year = 2016,
        month = feb,
       volume = {817},
       number = {2},
          eid = {111},
        pages = {111},
          doi = {10.3847/0004-637X/817/2/111},
archivePrefix = {arXiv},
       eprint = {1402.2655},
 primaryClass = {astro-ph.CO},
       adsurl = {https://ui.adsabs.harvard.edu/abs/2016ApJ...817..111D}
}

@ARTICLE{Murphy2019MNRAS.482.3458M,
       author = {{Murphy}, Michael T. and {Kacprzak}, Glenn G. and {Savorgnan}, Giulia A.~D. and {Carswell}, Robert F.},
        title = "{The UVES Spectral Quasar Absorption Database (SQUAD) data release 1: the first 10 million seconds}",
      journal = {\mnras},
         year = 2019,
        month = jan,
       volume = {482},
       number = {3},
        pages = {3458-3479},
          doi = {10.1093/mnras/sty2834},
archivePrefix = {arXiv},
       eprint = {1810.06136},
 primaryClass = {astro-ph.GA},
       adsurl = {https://ui.adsabs.harvard.edu/abs/2019MNRAS.482.3458M}
}

@ARTICLE{Weinberger2020ApJS..248...32W,
       author = {{Weinberger}, Rainer and {Springel}, Volker and {Pakmor}, R{\"u}diger},
        title = "{The AREPO Public Code Release}",
      journal = {\apjs},
         year = 2020,
        month = jun,
       volume = {248},
       number = {2},
          eid = {32},
        pages = {32},
          doi = {10.3847/1538-4365/ab908c},
archivePrefix = {arXiv},
       eprint = {1909.04667},
 primaryClass = {astro-ph.IM},
       adsurl = {https://ui.adsabs.harvard.edu/abs/2020ApJS..248...32W}
}

@ARTICLE{Faucher2020MNRAS.493.1614F,
       author = {{Faucher-Gigu{\`e}re}, Claude-Andr{\'e}},
        title = "{A cosmic UV/X-ray background model update}",
      journal = {\mnras},
         year = 2020,
        month = apr,
       volume = {493},
       number = {2},
        pages = {1614-1632},
          doi = {10.1093/mnras/staa302},
archivePrefix = {arXiv},
       eprint = {1903.08657},
 primaryClass = {astro-ph.CO},
       adsurl = {https://ui.adsabs.harvard.edu/abs/2020MNRAS.493.1614F}
}

@ARTICLE{Pillepich2018MNRAS.473.4077P,
       author = {{Pillepich}, Annalisa and {Springel}, Volker and {Nelson}, Dylan and {Genel}, Shy and {Naiman}, Jill and {Pakmor}, R{\"u}diger and {Hernquist}, Lars and {Torrey}, Paul and {Vogelsberger}, Mark and {Weinberger}, Rainer and {Marinacci}, Federico},
        title = "{Simulating galaxy formation with the IllustrisTNG model}",
      journal = {\mnras},
         year = 2018,
        month = jan,
       volume = {473},
       number = {3},
        pages = {4077-4106},
          doi = {10.1093/mnras/stx2656},
archivePrefix = {arXiv},
       eprint = {1703.02970},
 primaryClass = {astro-ph.GA},
       adsurl = {https://ui.adsabs.harvard.edu/abs/2018MNRAS.473.4077P}
}

@ARTICLE{Agarap2018arXiv180308375A,
       author = {{Agarap}, Abien Fred},
        title = "{Deep Learning using Rectified Linear Units (ReLU)}",
      journal = {arXiv e-prints},
         year = 2018,
        month = mar,
          eid = {arXiv:1803.08375},
        pages = {arXiv:1803.08375},
          doi = {10.48550/arXiv.1803.08375},
archivePrefix = {arXiv},
       eprint = {1803.08375},
 primaryClass = {cs.NE},
       adsurl = {https://ui.adsabs.harvard.edu/abs/2018arXiv180308375A}
}

@ARTICLE{Snoek2012arXiv1206.2944S,
       author = {{Snoek}, Jasper and {Larochelle}, Hugo and {Adams}, Ryan P.},
        title = "{Practical Bayesian Optimization of Machine Learning Algorithms}",
      journal = {arXiv e-prints},
         year = 2012,
        month = jun,
          eid = {arXiv:1206.2944},
        pages = {arXiv:1206.2944},
          doi = {10.48550/arXiv.1206.2944},
archivePrefix = {arXiv},
       eprint = {1206.2944},
 primaryClass = {stat.ML},
       adsurl = {https://ui.adsabs.harvard.edu/abs/2012arXiv1206.2944S}
}

@ARTICLE{Busca2013A&A...552A..96B,
       author = {{Busca}, N.~G. and {Delubac}, T. and {Rich}, J. and {Bailey}, S. and {Font-Ribera}, A. and {Kirkby}, D. and {Le Goff}, J.-M. and {Pieri}, M.~M. and {Slosar}, A. and {Aubourg}, {\'E}. and {Bautista}, J.~E. and {Bizyaev}, D. and {Blomqvist}, M. and {Bolton}, A.~S. and {Bovy}, J. and {Brewington}, H. and {Borde}, A. and {Brinkmann}, J. and {Carithers}, B. and {Croft}, R.~A.~C. and {Dawson}, K.~S. and {Ebelke}, G. and {Eisenstein}, D.~J. and {Hamilton}, J.-C. and {Ho}, S. and {Hogg}, D.~W. and {Honscheid}, K. and {Lee}, K.-G. and {Lundgren}, B. and {Malanushenko}, E. and {Malanushenko}, V. and {Margala}, D. and {Maraston}, C. and {Mehta}, K. and {Miralda-Escud{\'e}}, J. and {Myers}, A.~D. and {Nichol}, R.~C. and {Noterdaeme}, P. and {Olmstead}, M.~D. and {Oravetz}, D. and {Palanque-Delabrouille}, N. and {Pan}, K. and {P{\^a}ris}, I. and {Percival}, W.~J. and {Petitjean}, P. and {Roe}, N.~A. and {Rollinde}, E. and {Ross}, N.~P. and {Rossi}, G. and {Schlegel}, D.~J. and {Schneider}, D.~P. and {Shelden}, A. and {Sheldon}, E.~S. and {Simmons}, A. and {Snedden}, S. and {Tinker}, J.~L. and {Viel}, M. and {Weaver}, B.~A. and {Weinberg}, D.~H. and {White}, M. and {Y{\`e}che}, C. and {York}, D.~G.},
        title = "{Baryon acoustic oscillations in the Ly{\ensuremath{\alpha}} forest of BOSS quasars}",
      journal = {\aap},
         year = 2013,
        month = apr,
       volume = {552},
          eid = {A96},
        pages = {A96},
          doi = {10.1051/0004-6361/201220724},
archivePrefix = {arXiv},
       eprint = {1211.2616},
 primaryClass = {astro-ph.CO},
       adsurl = {https://ui.adsabs.harvard.edu/abs/2013A&A...552A..96B}
}

@ARTICLE{Slosar2013JCAP...04..026S,
       author = {{Slosar}, An{\v{z}}e and {Ir{\v{s}}i{\v{c}}}, Vid and {Kirkby}, David and {Bailey}, Stephen and {Busca}, Nicol{\'a}s G. and {Delubac}, Timoth{\'e}e and {Rich}, James and {Aubourg}, {\'E}ric and {Bautista}, Julian E. and {Bhardwaj}, Vaishali and {Blomqvist}, Michael and {Bolton}, Adam S. and {Bovy}, Jo and {Brownstein}, Joel and {Carithers}, Bill and {Croft}, Rupert A.~C. and {Dawson}, Kyle S. and {Font-Ribera}, Andreu and {Le Goff}, J.-M. and {Ho}, Shirley and {Honscheid}, Klaus and {Lee}, Khee-Gan and {Margala}, Daniel and {McDonald}, Patrick and {Medolin}, Bumbarija and {Miralda-Escud{\'e}}, Jordi and {Myers}, Adam D. and {Nichol}, Robert C. and {Noterdaeme}, Pasquier and {Palanque-Delabrouille}, Nathalie and {P{\^a}ris}, Isabelle and {Petitjean}, Patrick and {Pieri}, Matthew M. and {Pi{\v{s}}kur}, Yodovina and {Roe}, Natalie A. and {Ross}, Nicholas P. and {Rossi}, Graziano and {Schlegel}, David J. and {Schneider}, Donald P. and {Suzuki}, Nao and {Sheldon}, Erin S. and {Seljak}, Uro{\v{s}} and {Viel}, Matteo and {Weinberg}, David H. and {Y{\`e}che}, Christophe},
        title = "{Measurement of baryon acoustic oscillations in the Lyman-{\ensuremath{\alpha}} forest fluctuations in BOSS data release 9}",
      journal = {\jcap},
         year = 2013,
        month = apr,
       volume = {2013},
       number = {4},
          eid = {026},
        pages = {026},
          doi = {10.1088/1475-7516/2013/04/026},
archivePrefix = {arXiv},
       eprint = {1301.3459},
 primaryClass = {astro-ph.CO},
       adsurl = {https://ui.adsabs.harvard.edu/abs/2013JCAP...04..026S}
}

@ARTICLE{Bourboux2020ApJ...901..153D,
       author = {{du Mas des Bourboux}, H{\'e}lion and {Rich}, James and {Font-Ribera}, Andreu and {de Sainte Agathe}, Victoria and {Farr}, James and {Etourneau}, Thomas and {Le Goff}, Jean-Marc and {Cuceu}, Andrei and {Balland}, Christophe and {Bautista}, Julian E. and {Blomqvist}, Michael and {Brinkmann}, Jonathan and {Brownstein}, Joel R. and {Chabanier}, Sol{\`e}ne and {Chaussidon}, Edmond and {Dawson}, Kyle and {Gonz{\'a}lez-Morales}, Alma X. and {Guy}, Julien and {Lyke}, Brad W. and {de la Macorra}, Axel and {Mueller}, Eva-Maria and {Myers}, Adam D. and {Nitschelm}, Christian and {Mu{\~n}oz Guti{\'e}rrez}, Andrea and {Palanque-Delabrouille}, Nathalie and {Parker}, James and {Percival}, Will J. and {P{\'e}rez-R{\`a}fols}, Ignasi and {Petitjean}, Patrick and {Pieri}, Matthew M. and {Ravoux}, Corentin and {Rossi}, Graziano and {Schneider}, Donald P. and {Seo}, Hee-Jong and {Slosar}, An{\v{z}}e and {Stermer}, Julianna and {Vivek}, M. and {Y{\`e}che}, Christophe and {Youles}, Samantha},
        title = "{The Completed SDSS-IV Extended Baryon Oscillation Spectroscopic Survey: Baryon Acoustic Oscillations with Ly{\ensuremath{\alpha}} Forests}",
      journal = {\apj},
         year = 2020,
        month = oct,
       volume = {901},
       number = {2},
          eid = {153},
        pages = {153},
          doi = {10.3847/1538-4357/abb085},
archivePrefix = {arXiv},
       eprint = {2007.08995},
 primaryClass = {astro-ph.CO},
       adsurl = {https://ui.adsabs.harvard.edu/abs/2020ApJ...901..153D}
}

@ARTICLE{Faucher2009ApJ...703.1416F,
       author = {{Faucher-Gigu{\`e}re}, Claude-Andr{\'e} and {Lidz}, Adam and {Zaldarriaga}, Matias and {Hernquist}, Lars},
        title = "{A New Calculation of the Ionizing Background Spectrum and the Effects of He II Reionization}",
      journal = {\apj},
         year = 2009,
        month = oct,
       volume = {703},
       number = {2},
        pages = {1416-1443},
          doi = {10.1088/0004-637X/703/2/1416},
archivePrefix = {arXiv},
       eprint = {0901.4554},
 primaryClass = {astro-ph.CO},
       adsurl = {https://ui.adsabs.harvard.edu/abs/2009ApJ...703.1416F}
}

@ARTICLE{Haardt2012ApJ...746..125H,
       author = {{Haardt}, Francesco and {Madau}, Piero},
        title = "{Radiative Transfer in a Clumpy Universe. IV. New Synthesis Models of the Cosmic UV/X-Ray Background}",
      journal = {\apj},
         year = 2012,
        month = feb,
       volume = {746},
       number = {2},
          eid = {125},
        pages = {125},
          doi = {10.1088/0004-637X/746/2/125},
       adsurl = {https://ui.adsabs.harvard.edu/abs/2012ApJ...746..125H}
}

@ARTICLE{Crain2015MNRAS.450.1937C,
       author = {{Crain}, Robert A. and {Schaye}, Joop and {Bower}, Richard G. and {Furlong}, Michelle and {Schaller}, Matthieu and {Theuns}, Tom and {Dalla Vecchia}, Claudio and {Frenk}, Carlos S. and {McCarthy}, Ian G. and {Helly}, John C. and {Jenkins}, Adrian and {Rosas-Guevara}, Yetli M. and {White}, Simon D.~M. and {Trayford}, James W.},
        title = "{The EAGLE simulations of galaxy formation: calibration of subgrid physics and model variations}",
      journal = {\mnras},
         year = 2015,
        month = jun,
       volume = {450},
       number = {2},
        pages = {1937-1961},
          doi = {10.1093/mnras/stv725},
archivePrefix = {arXiv},
       eprint = {1501.01311},
 primaryClass = {astro-ph.GA},
       adsurl = {https://ui.adsabs.harvard.edu/abs/2015MNRAS.450.1937C}
}

@ARTICLE{Dave2019MNRAS.486.2827D,
       author = {{Dav{\'e}}, Romeel and {Angl{\'e}s-Alc{\'a}zar}, Daniel and {Narayanan}, Desika and {Li}, Qi and {Rafieferantsoa}, Mika H. and {Appleby}, Sarah},
        title = "{SIMBA: Cosmological simulations with black hole growth and feedback}",
      journal = {\mnras},
         year = 2019,
        month = jun,
       volume = {486},
       number = {2},
        pages = {2827-2849},
          doi = {10.1093/mnras/stz937},
archivePrefix = {arXiv},
       eprint = {1901.10203},
 primaryClass = {astro-ph.GA},
       adsurl = {https://ui.adsabs.harvard.edu/abs/2019MNRAS.486.2827D}
}

@ARTICLE{Almgren2013ApJ...765...39A,
       author = {{Almgren}, Ann S. and {Bell}, John B. and {Lijewski}, Mike J. and {Luki{\'c}}, Zarija and {Van Andel}, Ethan},
        title = "{Nyx: A Massively Parallel AMR Code for Computational Cosmology}",
      journal = {\apj},
         year = 2013,
        month = mar,
       volume = {765},
       number = {1},
          eid = {39},
        pages = {39},
          doi = {10.1088/0004-637X/765/1/39},
archivePrefix = {arXiv},
       eprint = {1301.4498},
 primaryClass = {astro-ph.IM},
       adsurl = {https://ui.adsabs.harvard.edu/abs/2013ApJ...765...39A}
}

@ARTICLE{Bolton2017MNRAS.464..897B,
       author = {{Bolton}, James S. and {Puchwein}, Ewald and {Sijacki}, Debora and {Haehnelt}, Martin G. and {Kim}, Tae-Sun and {Meiksin}, Avery and {Regan}, John A. and {Viel}, Matteo},
        title = "{The Sherwood simulation suite: overview and data comparisons with the Lyman {\ensuremath{\alpha}} forest at redshifts 2 {\ensuremath{\leq}} z {\ensuremath{\leq}} 5}",
      journal = {\mnras},
         year = 2017,
        month = jan,
       volume = {464},
       number = {1},
        pages = {897-914},
          doi = {10.1093/mnras/stw2397},
archivePrefix = {arXiv},
       eprint = {1605.03462},
 primaryClass = {astro-ph.CO},
       adsurl = {https://ui.adsabs.harvard.edu/abs/2017MNRAS.464..897B}
}

@ARTICLE{2024MNRAS.527.4545K,
       author = {{Khaire}, Vikram and {Hu}, Teng and {Hennawi}, Joseph F. and {Walther}, Michael and {Davies}, Frederick},
        title = "{Can the low-redshift Lyman alpha forest constrain AGN feedback models?}",
      journal = {\mnras},
         year = 2024,
        month = jan,
       volume = {527},
       number = {3},
        pages = {4545-4562},
          doi = {10.1093/mnras/stad3374},
archivePrefix = {arXiv},
       eprint = {2306.05466},
 primaryClass = {astro-ph.CO},
       adsurl = {https://ui.adsabs.harvard.edu/abs/2024MNRAS.527.4545K}
}

@ARTICLE{Hu2024MNRAS.52711338H,
       author = {{Hu}, Teng and {Khaire}, Vikram and {Hennawi}, Joseph F. and {O{\~n}orbe}, Jose and {Walther}, Michael and {Lukic}, Zarija and {Davies}, Frederick},
        title = "{The impact of the WHIM on the IGM thermal state determined from the low-z Lyman {\ensuremath{\alpha}} forest}",
      journal = {\mnras},
         year = 2024,
        month = feb,
       volume = {527},
       number = {4},
        pages = {11338-11359},
          doi = {10.1093/mnras/stad3846},
archivePrefix = {arXiv},
       eprint = {2308.14738},
 primaryClass = {astro-ph.CO},
       adsurl = {https://ui.adsabs.harvard.edu/abs/2024MNRAS.52711338H}
}

@ARTICLE{Gal2015arXiv150602142G,
       author = {{Gal}, Yarin and {Ghahramani}, Zoubin},
        title = "{Dropout as a Bayesian Approximation: Representing Model Uncertainty in Deep Learning}",
      journal = {arXiv e-prints},
         year = 2015,
        month = jun,
          eid = {arXiv:1506.02142},
        pages = {arXiv:1506.02142},
          doi = {10.48550/arXiv.1506.02142},
archivePrefix = {arXiv},
       eprint = {1506.02142},
 primaryClass = {stat.ML},
       adsurl = {https://ui.adsabs.harvard.edu/abs/2015arXiv150602142G}
}

@ARTICLE{Rudie2012ApJ...757L..30R,
       author = {{Rudie}, Gwen C. and {Steidel}, Charles C. and {Pettini}, Max},
        title = "{The Temperature-Density Relation in the Intergalactic Medium at Redshift langzrang = 2.4}",
      journal = {\apjl},
         year = 2012,
        month = oct,
       volume = {757},
       number = {2},
          eid = {L30},
        pages = {L30},
          doi = {10.1088/2041-8205/757/2/L30},
archivePrefix = {arXiv},
       eprint = {1209.0005},
 primaryClass = {astro-ph.CO},
       adsurl = {https://ui.adsabs.harvard.edu/abs/2012ApJ...757L..30R}
}

@ARTICLE{Gaikwad2017MNRAS.466..838G,
       author = {{Gaikwad}, Prakash and {Khaire}, Vikram and {Choudhury}, Tirthankar Roy and {Srianand}, Raghunathan},
        title = "{Intergalactic Lyman continuum photon budget in the past 5 billion years}",
      journal = {\mnras},
         year = 2017,
        month = apr,
       volume = {466},
       number = {1},
        pages = {838-860},
          doi = {10.1093/mnras/stw3086},
archivePrefix = {arXiv},
       eprint = {1605.02738},
 primaryClass = {astro-ph.CO},
       adsurl = {https://ui.adsabs.harvard.edu/abs/2017MNRAS.466..838G}
}

@ARTICLE{Gaikwad2020MNRAS.494.5091G,
       author = {{Gaikwad}, Prakash and {Rauch}, Michael and {Haehnelt}, Martin G. and {Puchwein}, Ewald and {Bolton}, James S. and {Keating}, Laura C. and {Kulkarni}, Girish and {Ir{\v{s}}i{\v{c}}}, Vid and {Ba{\~n}ados}, Eduardo and {Becker}, George D. and {Boera}, Elisa and {Zahedy}, Fakhri S. and {Chen}, Hsiao-Wen and {Carswell}, Robert F. and {Chardin}, Jonathan and {Rorai}, Alberto},
        title = "{Probing the thermal state of the intergalactic medium at z > 5 with the transmission spikes in high-resolution Ly {\ensuremath{\alpha}} forest spectra}",
      journal = {\mnras},
         year = 2020,
        month = jun,
       volume = {494},
       number = {4},
        pages = {5091-5109},
          doi = {10.1093/mnras/staa907},
archivePrefix = {arXiv},
       eprint = {2001.10018},
 primaryClass = {astro-ph.CO},
       adsurl = {https://ui.adsabs.harvard.edu/abs/2020MNRAS.494.5091G}
}

@ARTICLE{Gaikwad2021MNRAS.506.4389G,
       author = {{Gaikwad}, Prakash and {Srianand}, Raghunathan and {Haehnelt}, Martin G. and {Choudhury}, Tirthankar Roy},
        title = "{A consistent and robust measurement of the thermal state of the IGM at 2 {\ensuremath{\leq}} z {\ensuremath{\leq}} 4 from a large sample of Ly {\ensuremath{\alpha}} forest spectra: evidence for late and rapid He II reionization}",
      journal = {\mnras},
         year = 2021,
        month = sep,
       volume = {506},
       number = {3},
        pages = {4389-4412},
          doi = {10.1093/mnras/stab2017},
archivePrefix = {arXiv},
       eprint = {2009.00016},
 primaryClass = {astro-ph.CO},
       adsurl = {https://ui.adsabs.harvard.edu/abs/2021MNRAS.506.4389G}
}

@ARTICLE{Hu2026arXiv260605006H,
       author = {{Hu}, Teng and {Khaire}, Vikram and {Hennawi}, Joseph F. and {Tripp}, Todd M. and {Onorbe}, Jose and {Walther}, Michael and {Lukic}, Zarija},
        title = "{A Measurement of the Thermal and Ionization State of the IGM at $z < 0.5$}",
      journal = {arXiv e-prints},
         year = 2026,
        month = jun,
          eid = {arXiv:2606.05006},
        pages = {arXiv:2606.05006},
          doi = {10.48550/arXiv.2606.05006},
archivePrefix = {arXiv},
       eprint = {2606.05006},
 primaryClass = {astro-ph.CO},
       adsurl = {https://ui.adsabs.harvard.edu/abs/2026arXiv260605006H}
}

@ARTICLE{Bolton2008MNRAS.386.1131B,
       author = {{Bolton}, J.~S. and {Viel}, M. and {Kim}, T.-S. and {Haehnelt}, M.~G. and {Carswell}, R.~F.},
        title = "{Possible evidence for an inverted temperature-density relation in the intergalactic medium from the flux distribution of the Ly{\ensuremath{\alpha}} forest}",
      journal = {\mnras},
         year = 2008,
        month = may,
       volume = {386},
       number = {2},
        pages = {1131-1144},
          doi = {10.1111/j.1365-2966.2008.13114.x},
archivePrefix = {arXiv},
       eprint = {0711.2064},
 primaryClass = {astro-ph},
       adsurl = {https://ui.adsabs.harvard.edu/abs/2008MNRAS.386.1131B}
}

@ARTICLE{Becker2013MNRAS.436.1023B,
       author = {{Becker}, George D. and {Bolton}, James S.},
        title = "{New measurements of the ionizing ultraviolet background over 2 < z < 5 and implications for hydrogen reionization}",
      journal = {\mnras},
         year = 2013,
        month = dec,
       volume = {436},
       number = {2},
        pages = {1023-1039},
          doi = {10.1093/mnras/stt1610},
archivePrefix = {arXiv},
       eprint = {1307.2259},
 primaryClass = {astro-ph.CO},
       adsurl = {https://ui.adsabs.harvard.edu/abs/2013MNRAS.436.1023B}
}

@ARTICLE{Khaire2019MNRAS.484.4174K,
       author = {{Khaire}, Vikram and {Srianand}, Raghunathan},
        title = "{New synthesis models of consistent extragalactic background light over cosmic time}",
      journal = {\mnras},
         year = 2019,
        month = apr,
       volume = {484},
       number = {3},
        pages = {4174-4199},
          doi = {10.1093/mnras/stz174},
archivePrefix = {arXiv},
       eprint = {1801.09693},
 primaryClass = {astro-ph.GA},
       adsurl = {https://ui.adsabs.harvard.edu/abs/2019MNRAS.484.4174K}
}

@ARTICLE{Khaire2015ApJ...805...33K,
       author = {{Khaire}, Vikram and {Srianand}, Raghunathan},
        title = "{Star Formation History, Dust Attenuation, and Extragalactic Background Light}",
      journal = {\apj},
         year = 2015,
        month = may,
       volume = {805},
       number = {1},
          eid = {33},
        pages = {33},
          doi = {10.1088/0004-637X/805/1/33},
archivePrefix = {arXiv},
       eprint = {1405.7038},
 primaryClass = {astro-ph.GA},
       adsurl = {https://ui.adsabs.harvard.edu/abs/2015ApJ...805...33K}
}

@ARTICLE{Puchwein2019MNRAS.485...47P,
       author = {{Puchwein}, Ewald and {Haardt}, Francesco and {Haehnelt}, Martin G. and {Madau}, Piero},
        title = "{Consistent modelling of the meta-galactic UV background and the thermal/ionization history of the intergalactic medium}",
      journal = {\mnras},
         year = 2019,
        month = may,
       volume = {485},
       number = {1},
        pages = {47-68},
          doi = {10.1093/mnras/stz222},
archivePrefix = {arXiv},
       eprint = {1801.04931},
 primaryClass = {astro-ph.GA},
       adsurl = {https://ui.adsabs.harvard.edu/abs/2019MNRAS.485...47P}
}

@ARTICLE{Hu2025MNRAS.536....1H,
       author = {{Hu}, Teng and {Khaire}, Vikram and {Hennawi}, Joseph F. and {Tripp}, Todd M. and {O{\~n}orbe}, Jose and {Walther}, Michael and {Luki{\'c}}, Zarija},
        title = "{Measurements of the thermal and ionization state of the intergalactic medium during the cosmic afternoon}",
      journal = {\mnras},
         year = 2025,
        month = jan,
       volume = {536},
       number = {1},
        pages = {1-26},
          doi = {10.1093/mnras/stae2474},
archivePrefix = {arXiv},
       eprint = {2311.17895},
 primaryClass = {astro-ph.CO},
       adsurl = {https://ui.adsabs.harvard.edu/abs/2025MNRAS.536....1H}
}

@ARTICLE{Alam2021PhRvD.103h3533A,
       author = {{Alam}, Shadab and {Aubert}, Marie and {Avila}, Santiago and {Balland}, Christophe and {Bautista}, Julian E. and {Bershady}, Matthew A. and {Bizyaev}, Dmitry and {Blanton}, Michael R. and {Bolton}, Adam S. and {Bovy}, Jo and {Brinkmann}, Jonathan and {Brownstein}, Joel R. and {Burtin}, Etienne and {Chabanier}, Sol{\`e}ne and {Chapman}, Michael J. and {Choi}, Peter Doohyun and {Chuang}, Chia-Hsun and {Comparat}, Johan and {Cousinou}, Marie-Claude and {Cuceu}, Andrei and {Dawson}, Kyle S. and {de la Torre}, Sylvain and {de Mattia}, Arnaud and {Agathe}, Victoria de Sainte and {des Bourboux}, H{\'e}lion du Mas and {Escoffier}, Stephanie and {Etourneau}, Thomas and {Farr}, James and {Font-Ribera}, Andreu and {Frinchaboy}, Peter M. and {Fromenteau}, Sebastien and {Gil-Mar{\'\i}n}, H{\'e}ctor and {Le Goff}, Jean-Marc and {Gonzalez-Morales}, Alma X. and {Gonzalez-Perez}, Violeta and {Grabowski}, Kathleen and {Guy}, Julien and {Hawken}, Adam J. and {Hou}, Jiamin and {Kong}, Hui and {Parker}, James and {Klaene}, Mark and {Kneib}, Jean-Paul and {Lin}, Sicheng and {Long}, Daniel and {Lyke}, Brad W. and {de la Macorra}, Axel and {Martini}, Paul and {Masters}, Karen and {Mohammad}, Faizan G. and {Moon}, Jeongin and {Mueller}, Eva-Maria and {Mu{\~n}oz-Guti{\'e}rrez}, Andrea and {Myers}, Adam D. and {Nadathur}, Seshadri and {Neveux}, Richard and {Newman}, Jeffrey A. and {Noterdaeme}, Pasquier and {Oravetz}, Audrey and {Oravetz}, Daniel and {Palanque-Delabrouille}, Nathalie and {Pan}, Kaike and {Paviot}, Romain and {Percival}, Will J. and {P{\'e}rez-R{\`a}fols}, Ignasi and {Petitjean}, Patrick and {Pieri}, Matthew M. and {Prakash}, Abhishek and {Raichoor}, Anand and {Ravoux}, Corentin and {Rezaie}, Mehdi and {Rich}, James and {Ross}, Ashley J. and {Rossi}, Graziano and {Ruggeri}, Rossana and {Ruhlmann-Kleider}, Vanina and {S{\'a}nchez}, Ariel G. and {S{\'a}nchez}, F. Javier and {S{\'a}nchez-Gallego}, Jos{\'e} R. and {Sayres}, Conor and {Schneider}, Donald P. and {Seo}, Hee-Jong and {Shafieloo}, Arman and {Slosar}, An{\v{z}}e and {Smith}, Alex and {Stermer}, Julianna and {Tamone}, Amelie and {Tinker}, Jeremy L. and {Tojeiro}, Rita and {Vargas-Maga{\~n}a}, Mariana and {Variu}, Andrei and {Wang}, Yuting and {Weaver}, Benjamin A. and {Weijmans}, Anne-Marie and {Y{\`e}che}, Christophe and {Zarrouk}, Pauline and {Zhao}, Cheng and {Zhao}, Gong-Bo and {Zheng}, Zheng},
        title = "{Completed SDSS-IV extended Baryon Oscillation Spectroscopic Survey: Cosmological implications from two decades of spectroscopic surveys at the Apache Point Observatory}",
      journal = {\prd},
         year = 2021,
        month = apr,
       volume = {103},
       number = {8},
          eid = {083533},
        pages = {083533},
          doi = {10.1103/PhysRevD.103.083533},
archivePrefix = {arXiv},
       eprint = {2007.08991},
 primaryClass = {astro-ph.CO},
       adsurl = {https://ui.adsabs.harvard.edu/abs/2021PhRvD.103h3533A}
}

@ARTICLE{Khaire2019MNRAS.486..769K,
       author = {{Khaire}, Vikram and {Walther}, Michael and {Hennawi}, Joseph F. and {O{\~n}orbe}, Jose and {Luki{\'c}} and {}, Zarija and {Prochaska}, J. Xavier and {Tripp}, Todd M. and {Burchett}, Joseph N. and {Rodriguez}, Christian},
        title = "{The power spectrum of the Lyman-{\ensuremath{\alpha}} Forest at z < 0.5}",
      journal = {\mnras},
         year = 2019,
        month = jun,
       volume = {486},
       number = {1},
        pages = {769-782},
          doi = {10.1093/mnras/stz344},
archivePrefix = {arXiv},
       eprint = {1808.05605},
 primaryClass = {astro-ph.CO},
       adsurl = {https://ui.adsabs.harvard.edu/abs/2019MNRAS.486..769K}
}

@ARTICLE{Kim2013A&A...552A..77K,
       author = {{Kim}, T.-S. and {Partl}, A.~M. and {Carswell}, R.~F. and {M{\"u}ller}, V.},
        title = "{The evolution of H I and C IV quasar absorption line systems at 1.9 < z < 3.2}",
      journal = {\aap},
         year = 2013,
        month = apr,
       volume = {552},
          eid = {A77},
        pages = {A77},
          doi = {10.1051/0004-6361/201220042},
archivePrefix = {arXiv},
       eprint = {1302.6622},
 primaryClass = {astro-ph.CO},
       adsurl = {https://ui.adsabs.harvard.edu/abs/2013A&A...552A..77K}
}








\bsp	
\label{lastpage}
\end{document}